\documentclass[sigconf, nonacm]{acmart}

\usepackage[ruled, vlined, linesnumbered]{algorithm2e}
\usepackage{algorithmic}
\usepackage{subfig}
\usepackage{epsfig}
\newcommand{\graph}{\ensuremath{\mathcal{G}}}
\newcommand{\edges}{\ensuremath{\mathcal{E}}}
\newcommand{\vertices}{\ensuremath{\mathcal{V}}}
\newcommand{\elabel}{\ensuremath{\ell}}
\newcommand{\elabels}{\ensuremath{\mathcal{L}}}
\newcommand{\attributes}{\ensuremath{\mathcal{A}}}
\newcommand{\rules}{\ensuremath{\mathcal{R}}}

\newcommand{\pathpatterns}{\ensuremath{\mathcal{P}}}
\newcommand{\minsup}{\ensuremath{\theta}}
\newcommand{\minlen}{\ensuremath{k}}

\newcommand{\candidates}{\ensuremath{\mathcal{C}}}

\newcommand{\labelfont}[1]{\texttt{#1}}

\setcopyright{acmcopyright}
\copyrightyear{2018}
\acmYear{2018}
\acmDOI{XXXXXXX.XXXXXXX}

\acmConference[Conference acronym 'XX]{Make sure to enter the correct
  conference title from your rights confirmation emai}{June 03--05,
  2018}{Woodstock, NY}
\acmPrice{15.00}
\acmISBN{978-1-4503-XXXX-X/18/06}

\begin{document}

\title{Path Association Rule Mining}

\author{Yuya Sasaki}
\email{sasaki@ist.osaka-u.ac.jp}
\affiliation{%
  \institution{Osaka University}
}

\begin{abstract}
Graph association rule mining is a data mining technique used for discovering regularities in graph data. 
In this study, we propose a novel concept, {\it path association rule mining}, to discover the correlations of path patterns that frequently appear in a given graph.
Reachability path patterns (i.e., existence of paths from a vertex to another vertex) are applied in our concept to discover diverse regularities.
We show that the problem is NP-hard, and we develop an efficient algorithm in which the anti-monotonic property is used on path patterns. 
Subsequently, we develop approximation and parallelization techniques to efficiently and scalably discover rules.
We use real-life graphs to experimentally verify the effectiveness of path association rules and the efficiency of our algorithm. 
\end{abstract}

\maketitle

\section{Introduction}
\label{sec:introduction}

Association rule mining is data mining technique used for discovering regularities between itemsets in large databases~\cite{agrawal1993mining,zhao2003association}. 
Association rules are represented as $X \Rightarrow Y$, where $X$ and $Y$ are the disjoint and called antecedent and consequent, respectively.
Association rules have attracted research attention because of their usefulness in many applications such as market analysis~\cite{kaur2016market}, web mining~\cite{lee2001web}, and bioinformatics~\cite{mallik2014ranwar}.

Regularities between entities in large-scale graphs are a critical topic of research.
{\it Graph association rule mining} on a single large graph, which is an extended form of general association rule mining, exhibits considerable potential for studying regularities~\cite{fan2015association,wang2020extending,fan2016adding,fan2022discovering}.
Graph association rules are represented as $G_X \Rightarrow G_Y$, where $G_X$ and $G_Y$ are graph patterns.
Since graphs are widely used in many applications, association rules on graphs have a large opportunity to discovery valuable insights and knowledge.

\smallskip
\noindent
{\bf Applications on graph association rule mining.}
Graph association rule mining exhibits numerous application scenarios.

\noindent
\underline{Social analysis}:
Social analysis is critical for understanding and improving our life.
Social relationships can be modeled using graphs, and graph association rule mining can be used to discover regularities between the status of people and their relationship patterns. 
For example, social relationships affect to our health \cite{house1988social} and happiness \cite{haller2006social}.
We can find interesting rules, for example, ``if people are happy, they are healthy and have friends.''

\noindent
\underline{Recommendation}:
In current recommendation systems, user behavior, such as viewing and clicking \cite{lin2002efficient,davidson2010youtube}, toward particular products are tracked.
Bipartite graphs are used to model user behavior to products.
Graph association rule mining can be used find potential customers by finding regularities among user behavior, products, and profiles.

\noindent
\underline{Social discrimination checking}:
Social discrimination in graphs should be solved for building fair machine learning models \cite{bias2018}.
Machine learning models for graph data can be suffered from such discriminatory bias as same as transactional data.
Since graph association rules can discover regularities that include discriminatory bias, they can be used to remove discriminatory bias in the graph data.





\smallskip
\noindent
{\bf Motivation and challenge.}
Graph association rule mining is fundamental for graph analysis, but only a few techniques have been developed~\cite{fan2015association,wang2020extending,fan2016adding,fan2022discovering}.
Existing methods cannot be applied to the aforementioned applications due to their own semantics (see Sec. \ref{sec:related} in detail).

Existing works have four obstacles to apply the aforementioned applications.
First, in all existing works, a set of vertices in consequent (i.e., $G_Y$) must be a subset of those in the antecedent (i.e., $G_X$).
Thus, they cannot discover association rules such that $G_X$ and $G_Y$ involve distinct vertices.
Second, since $G_X$ and $G_Y$ are restrictive (e.g., a single edge or no common edges in $G_X$ and $G_Y$), they cannot find the difference of vertex attributes.
Third, each rule and discovery algorithm assumes a single graph type, for example, graphs with labeled vertices and unlabeled edges, and thus does not handle more generalized graphs than that it assumes (e.g., if discovery algorithms assume graphs with unlabeled edges, it cannot handle ones with labeled edges). 
Fourth, their antecedent and consequent are basic graph pattern, so it does not capture the connectivity between vertices.

Therefore, we need a novel graph association rule mining technique.
We introduce three challenges that have not been addressed in existing studies. 
First, simple association rules are required to flexibly capture regularities that can represent (1) no restriction of vertices in antecedent and consequent and (2) connectivity between vertices.
Second, an efficient and scalable mining algorithm is required to accurately find all frequent patterns in general property graphs that include vertex attributes and edge labels.
Finally, such rules should be effective and provide insight and knowledge, and offer conventional metrics such as support, confidence, and lift. 


\begin{figure*}[t]
    \centering
	\includegraphics[width=1.0\linewidth]{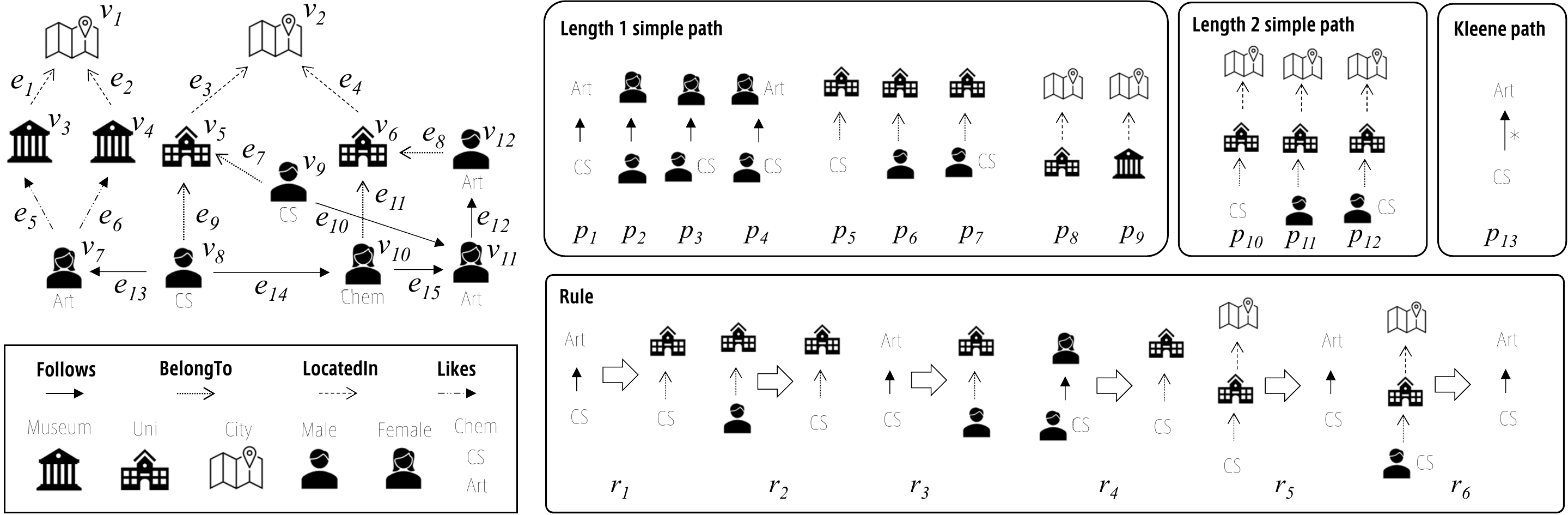}
	   \vspace{-6mm}
    \caption{Path association rules}
    \label{fig:intro}
\end{figure*}

\smallskip
\noindent
{\bf Contribution.}
We propose a novel and simple concept, {\it path association rule mining}, and develop an efficient and scalable algorithm for discovering path association rules. 
Path association rule mining aims to discover regularities between path patterns; path patterns are sequences of vertex attribute sets and edge labels.
{\it Reachablity} patterns can be applied as path patterns that represent label-constrained reachability between two vertices in directed edges.
This mining finds rules $p_X \Rightarrow p_Y$ such that more than $\minsup$ vertices become sources of path patterns $p_X$ and $p_Y$, where $\minsup$ is a given minimum support.
This rule can naturally offer standard metrics, such as absolute/relative support, confidence, and lift. 

\begin{example}
Figure~\ref{fig:intro} illustrates an example of path association rule mining.
A given graph includes 12 vertices, 15 edges, 4 types of edge labels, and 8 types of vertex attributes.
From this graph, we mine path association rules with $\minsup$ is one.
Path pattern $\langle \{ \labelfont{Male} \}, \labelfont{Follows}, \{\labelfont{Art} \} \rangle$ is frequent because vertices $v_8$ and $v_9$ become the sources of the path patterns (we say that $v_8$ and $v_9$ match with the path pattern).
Similarly, path pattern $\langle \{\labelfont{CS} \}, \labelfont{BelongTo}, \{\labelfont{Uni} \} \rangle$ matches $v_8$ and $v_9$.
$\langle \{\labelfont{CS} \}, \labelfont{Follows}, \{\labelfont{Art} \} \rangle$ $\Rightarrow$ $\langle \{\labelfont{Male} \}, \labelfont{BelongTo},$ $\{\labelfont{Uni} \} \rangle$ is discovered as a path association rule.
\end{example}

The path association rule problem is NP-hard. 
Therefore, we develop an algorithm to efficiently find the set of frequent path patterns and rules. 
Our algorithm employs an Apriori approach using anti-monotonic property that path association rules hold to prune the candidate of frequent path patterns and rules.
We can prune path patterns in two ways, namely vertically, which stops extending the length of path patterns, and horizontally, which stops extending the number of attributes in path patterns.
Our algorithm can exactly compute support, confidence, and other measures of rules.
Furthermore, we develop parallelization and approximation techniques to scalably accelerate finding association rules.
In our parallelization technique, we estimate costs of vertices to compute path patterns, and assign the set of vertices into threads according to the costs.
We reduce the computation costs in two approximate methods; reducing candidates of suffix of path patterns and sampling vertices to be computed. 


The contributions of the study are summarized as follows:

\begin{itemize}
    \item {\bf Novel Concept}: We propose path association rule mining to capture regularities between path patterns. We can apply reachability path patterns, which differs from existing graph association rule mining.
    \item {\bf Efficient and Scalable Algorithm}: We develop an efficient and scalable algorithm that parallely finds the rules with pruning infrequent path patterns. Our algorithm can apply approximation techniques for more efficiency.
    \item {\bf Discoveries in Real-World Graphs}: We demonstrate that path association rule mining is effective in knowledge extraction and bias checking.
\end{itemize}

\noindent
{\bf Reproducibility}. The code and datasets used in this study are available at github.
{\it Since codes on the existing works~\cite{fan2017big,fan2015association,wang2020extending,fan2022discovering,fan2016adding} are not open publicly, our code is the first open software on graph association rule mining on a single large graph.}
Our {\it supplementary file} in the github includes proofs of our theorems and lemmas, and additional related studies and experimental results.




\section{Our Concept}
\label{sec:concept}

We propose a novel concept, path association rule mining to effectively determine the regulations between attributes of vertices connected by edge labels.
Its remarkable feature compared with existing graph association rules is that it captures correlations of distinct path patterns from the same vertices, which are often useful in many applications.
In addition, the path association rule mining discovers rules on general property graphs, which widely cover many graph types (e.g., labeled graphs). 



\subsection{Notations}

We consider graph $\graph=(\vertices,\edges,\elabels,\attributes)$ where $\vertices$ is a finite set of vertices, $\edges \subset \vertices \times \elabels \times \vertices$ is the set of edges with label $\elabel \in \elabels$, $\elabels$ is a finite set of edge labels, and $\attributes$ is a finite set of attributes. 
Edge $e \in \edges$ is represented by a triple $(v, \elabel_e, v')$ where an edge from vertex $v$ to vertex $v'$ with edge label $\elabel_e$.
Attribute $a \in \attributes$ is a categorical value that represents a feature of vertex.
We define $A \subseteq \attributes$ as a set of attributes and each $v \in \vertices$ has a set of attributes $A(v)$.

{\it Path} is a sequence of vertices and edges $\langle v_0, e_0, v_1, \ldots, e_{n-1}, v_n \rangle$ where $n$ is a length of the path.
We define $v_0$ and $v_n$ as the source and target of path, respectively.

\begin{example}
In Figure~\ref{fig:intro}, $\elabels =\{ \labelfont{Follows}, \labelfont{BelongTo},$ $\labelfont{LocatedIn},$ $\labelfont{Likes} \}$ and $\attributes = \{ \labelfont{Museum}, \labelfont{Uni}, \labelfont{City}, \labelfont{Male}, \labelfont{Female},$ $\labelfont{CS}, \labelfont{Chem}, \labelfont{Art} \}$.
$\langle v_8, $ $e_9, v_5, e_3, v_2 \rangle$ is a two length path where $v_8$ and $v_2$ are the source and target of the path, respectively.
\end{example}

\subsection{Path Association Rule}

We here define path association rules that we propose in this paper, after defining path patterns.

\smallskip
\noindent
{\bf Path pattern: }
We define two path patterns, namely simple and reachability path patterns.

\begin{itemize}
\item {\it Simple path pattern} is a sequence of attribute sets and edge labels $\langle A_0, \elabel_0, A_1, \ldots, \elabel_{n-1}, A_n \rangle$ where $A_i \!\subseteq \!\attributes$ and $\elabel_i \!\in\! \elabels$.
Given vertex $v$ and simple path pattern $p$, $v$ matches $p$ if $v$ is a source of a path that satisfies $A_i \!\! \subseteq \!\! A(v_i)$ and $\elabel_i \!\! = \!\!\elabel_{e_i}$ for all $i$.

\item {\it Reachability path pattern} is a pair of attribute sets and edge label $\langle A_0, \elabel^*, A_1 \rangle$, where $A_0, A_1  \subseteq \attributes$ and $\elabel^* \in \elabels$.
Given vertex $v$ and reachability path pattern $p$, $v$ matches $p$ if $v$ is a source of a path that satisfies $A_0 \subseteq A(v_0)$, $A_1 \subseteq A(v_n)$, and $\elabel^* \!\! = \!\!\elabel_{e_i}$ for all $i$.
\end{itemize}

We denote by $\vertices(p)$ the set of all matched vertices of path pattern $p$ and by $|\vertices(p)|$ the number of matched vertices of $p$.
We call a path pattern whose all attributes sets include a single attribute {\it unit path pattern}.

\begin{definition}[Dominance]
Given two path patterns $p=\langle A_0, \elabel_0, A_1,$ $\ldots, \elabel_{n-1}, A_n \rangle$ and $p'=\langle A_0',$ $\elabel_0', A_1',$ $\ldots, \elabel_{m-1}', A_m' \rangle$, $p$ dominates $p'$  if $m \leq n$, and $\elabel_i' = \elabel_i$ and $A_i' \subseteq A_i$ for all $i = 0$ to $m$. We denote $p' \subset p$ when $p'$ is dominated by $p$.
\end{definition}

Intuitively, dominating path patterns are more complex than their dominated path patterns.

\begin{example}
In Figure~\ref{fig:intro}, $v_H$ matches simple path patterns $\langle \{\labelfont{CS},$ $\labelfont{Male} \}, \labelfont{Follows}, \{\labelfont{CS}, \labelfont{female} \} \rangle$ and reachability path patterns $\langle \{\labelfont{CS}, \labelfont{Male} \},$ $\labelfont{Follows}^{*}, \{\labelfont{CS}, \labelfont{Male} \} \rangle$.
$\vertices(\langle \{ \labelfont{Male} \}, \labelfont{Follows}, \{ \labelfont{Male} \} \rangle) $ is $\{v_8, v_9, v_{12} \}$.
\end{example}

\smallskip
\noindent
{\bf Path association rule: } We define path association rules as follows.

\begin{definition}
A path association rule $r$ is defined by $p_X \Rightarrow p_Y$, where $p_X$ and $p_Y$ are path patterns.
We refer to $p_X$ and $p_Y$ as the antecedent and consequent of rule, respectively.
\end{definition}

The path association rule evaluates frequencies and confidential probabilities of vertices that match both $p_X$ and $p_Y$.
We apply {\it homorphily semantics} on path association rules. Thus, we allow that a single path is shared by both $p_X$ and $p_Y$. 

\begin{example}
Given a rule $\langle \{\labelfont{CS} \}, \labelfont{Follows}, \{\labelfont{Art}\} \rangle \Rightarrow \langle \{\labelfont{CS} \},$ $\labelfont{BelongTo}, \{\labelfont{Uni}\} \rangle$ in Figure~\ref{fig:intro}, $v_8$ and $v_9$ are matched with this rule.
\end{example}




\subsection{Metrics of association rules}
The path association rule offers general metrics of association rules. 
We can naturally define support, confidence, and lift for path association rules.

\noindent
{\it Support: }
The support of path association rules indicates how many vertices are applied into sources of paths.
We define two support metrics of absolute and relative supports. 
Most of graph association rule mining does not offer relative support (see Sec~\ref{sec:related}) because it is hard to compute the possible maximum number of matched graph patterns.
The absolute support is defined as follows:
\begin{equation*}
    \mathit{ASupp}(p_X \Rightarrow p_Y) = |\vertices(p_X) \cap \vertices(p_Y)|.
\end{equation*}

Since the maximum value of $|\vertices(p)|$ is the number of vertices, the relative support is defined as follows:
\begin{equation*}
    \mathit{RSupp}(p_X\Rightarrow p_Y) = \frac{|\vertices(p_X) \cap \vertices(p_Y)|}{|\vertices|}.
\end{equation*}

\noindent
{\it Confidence: }
The confidence of path association rules indicates the probability that vertices satisfies $p_X$ if they satisfy $p_Y$.
The confidence is defined as follows:
\begin{equation*}
    \mathit{Conf}(p_X \Rightarrow p_Y) = \frac{|\vertices(p_X) \cap \vertices(p_Y)|}{|\vertices(p_X)|}.
\end{equation*}

\smallskip
\noindent
{\it Lift: } Lift is not offered by most of graph association rule mining as well as the relative support. We can define lift as follows:

\begin{equation*}
    \mathit{Lift}(p_X \Rightarrow p_Y) = \frac{ |\vertices(p_X) \cap \vertices(p_Y)|\cdot |\vertices|}{|\vertices(p_X)|\cdot|\vertices(p_Y)|}.
\end{equation*}

\smallskip
\noindent
{\it Others:} Similarly, we can define other measures.

\begin{example}
We explain the metrics by using Figure~\ref{fig:intro}.
$\mathit{ASupp(r_1)}$, $\mathit{RSupp(r_1)}$, $\mathit{Conf(r_1)}$, and $\mathit{Lift(r_1)}$ are 2, 0.16, 1.0, and 6, respectively.
\end{example}

\subsection{Problem Definition}

We define a problem that we solve in this paper.

\begin{definition}[Path association rule mining problem]
Given graph \graph, minimum (absolute or relative) support \minsup, maximum path length \minlen, the path association rule mining problem aims to find all rules with their metrics, where $r$ satisfies (1) $\mathit{supp}(r)$ is larger than \minsup, (2) lengths of paths are at most \minlen,  and (3) $p_X$ is not dominated $p_Y$ and vice versa.
We say that $p$ is frequent if  $|\vertices(p)| > \minsup$.
\end{definition}

\begin{theorem}
Path association rule mining problem is NP-hard.
\end{theorem}
\noindent
{\it Proof sketch}: We show that the problem is NP-hard through reduction from as frequent subsequence enumeration problem, which is NP-hard~\cite{yang2004complexity}.
Sequences and subsequences are sub classes of graphs and path patterns, respectively.
Thus, the path association rule mining problem is NP-hard. \hfill{} $\square$

\smallskip
\noindent
{\bf Remarks1: }
The path association rule is general compared with conventional association rule mining for itemsets.
Each vertex and its attributes can be considered as a transaction and items of the transaction, respectively.
If no edges exist in graphs, path association rule mining is equivalent to conventional association rule mining for itemsets. 

\smallskip
\noindent
{\bf Remarks2: }
In recent real-life query logs for Wikidata~\cite{vrandevcic2014wikidata}, more than 90 \% of queries are path patterns~\cite{bonifati2020analytical,BonifatiMT19}.
From this fact, graph analysis often utilizes path patterns instead of graph patterns to understand real-world relationships.
Therefore, path association rule mining can be useful in real-world analysis on many applications.
\section{Our Algorithm}
\label{sec:algorithm}

We present an efficient algorithm for path association rule mining.



\subsection{Overview}
For solving the path association rule mining problem, we need to enumerate frequent path patterns and vertices that match the path patterns.
To reduce the candidates of path patterns, our algorithm employs an Apriori strategy which initially finds non-complex frequent path patterns and generates candidates of frequent complex path patterns form the found path patterns, based on anti-monotonic properties.

\smallskip
\noindent
{\bf Anti-monotonic properties.}
Anti-monotonic properties are leveraged to reduce candidates of path patterns effectively.
Path patterns naturally satisfy anti-monotonic properties as follows:

\begin{theorem}\label{theo:dominate}
If $p' \subset p$, $\vertices(p') \supseteq \vertices(p)$.
\end{theorem}
{\it Proof}: When $p$ is dominated by $p'$, $p$ is not longer than $p'$, all edge labels on $p$ and $p'$ are the same order, and attribute sets on $p$ are subset or equal to those on $p'$.
All vertices that match $p'$ are matched with path patterns whose attributes are subset of $p'$.
Therefore, if path pattern $p \subseteq p'$, $\vertices(p) \supseteq \vertices(p')$.
\hfill{} $\square$

\noindent
We can derive two lemmas from Theorem~\ref{theo:dominate}. 
\begin{lemma}\label{lemma:antimonotonicity_1}
If prefix of path patterns is not frequent, the whole path patterns are not frequent.
\end{lemma}
{\it Proof}: If $v$ matches $p=(A_0, \elabel_1, \cdots, \elabel_{n}, A_m)$, $v$ matches $(A_0, \elabel_1, \cdots,$ $\elabel_{n}, A_m)$ where $m < n$.
Similarly, if  $v$ does note match $(A_0, \elabel_1, \cdots,$ $\elabel_{m}, A_m)$, $v$ does not matches $(A_0, \elabel_1, \cdots, \elabel_{n}, A_n)$.
Therefore, if prefix of path patterns is not frequent, the whole path patterns are not frequent.
\hfill{} $\square$

\begin{lemma}\label{lemma:antimonotonicity_2}
If $p$ is not frequent, $p'$ such that $\exists A_i' \supset A_i$ and $\elabel_i' = \elabel_i$ for all $i$ is not frequent.
\end{lemma}
{\it Proof}: If $v$ matches $p'$, $v$ matches $p$ because $p'$ is more complex than $p$.
Similarly, if  $v$ does note match $p$, $v$ does not matches $p'$.
Therefore, if $p$ is not frequent, $p'$ is not frequent.
\hfill{} $\square$

\noindent
These theorem and lemmas can be used for pruning path patterns that are not frequent. 

\begin{example}
We show anti-monotonic properties by using Figure~\ref{fig:intro}.
We assume the minimum absolute support is $1$.
Path pattern $p = \langle \{ \labelfont{CS} \}, \labelfont{Follows}, \{\labelfont{Chem} \} \rangle$ is not frequent, so no path patterns that extend the path pattern are frequent.
Similarly, path patterns that dominate $p$ such as $\langle \{ \labelfont{Male, CS} \}, \labelfont{Follows}, \{\labelfont{Female,Chem} \} \rangle$ are not frequent.
\end{example}

\noindent
{\bf Baseline algorithm.}
Our algorithm has the following steps: (1) Frequent attribute set discovery, (2) Frequent simple path pattern discovery, (3) Frequent reachability path pattern discovery, and (4) Rule discovery, based on Lemma~\ref{lemma:antimonotonicity_1}.
We describe these steps.

\noindent
{\bf (1) Frequent attribute set discovery.} 
We obtain a set $\pathpatterns_0$ of frequent attribute sets (i.e., frequent path patterns with path length zero). 
This step can employ any type of algorithms for conventional association rule mining, such as Apriori method.


\noindent
{\bf (2) Frequent simple path pattern discovery.}
Our algorithm finds simple path patterns by extending the found frequent path patterns.
It generates candidates of simple path patterns with length $i$ by combining $p \in \pathpatterns_{i-1}$, an edge label, and $A \subseteq \attributes$, and then checks their frequency. 
If the path patterns are frequent, our algorithm stores them as frequent simple path patterns into $\pathpatterns_{i}$.

\noindent
{\bf (3) Frequent reachability path pattern discovery}
This step finds vertices that match reachability path patterns.
Different from frequent simple path pattern discovery, this step needs a set of vertices that are reached through $\elabel^*$ from each vertex. 

Given an edge label, we enumerate reachable vertices from each vertex that has frequent attributes by breadth-first search.
Similarly to simple pattern discovery, we generate candidates of reachability path patterns by combining $p \in \pathpatterns_{0}$, an edge label, and $A \subseteq \attributes$, and then we store frequent ones to $\pathpatterns^*$. 

\noindent
{\bf (4) Rule discovery.} 
In the rule discovery step, we search for vertices that match two path patterns.
If $p$ and $p'$ are frequent path patterns found in steps (2) and (3), it generates candidates of rules $p \Rightarrow p'$ and $p' \Rightarrow p$ and counts vertices that match both $p$ and $p'$.
Finally, it computes the metrics of rules if they are frequent.

\subsection{Optimization Strategy}

We enhance the efficiency of our algorithm by reducing the candidates of frequent path patterns and rules. 
We introduce suffix pruning, and then we explain how to enhance the candidate generations in our algorithm.

\smallskip
\noindent
{\bf Suffix pruning.}
Anti-monotonic properties reduce a prefix of path patterns. However, they do not reduce a suffix.
Therefore, we introduce a suffix pruning.

Given given edge label $\elabel$ and attribute $a$, we compute the maximum number of matched vertices for any path patterns of length $n-1$ concatenated with $\elabel$ and $A$.
In addition, given $A$ and $\elabel$, we compute the maximum number of matched vertices for $\langle A, \elabel, \ldots \rangle$.
We can prune the suffix if the maximum number of vertices is not larger than $\minsup$. 

We define two values. First, the number of vertices that have $A$ and out-going edges with edge label $\elabel$ is $|\vertices(A, \elabel)| = |\{v| (v,\elabel_e,v') \in \edges \land A(v) \supseteq A \land \elabel_e = \elabel\}|$.
Second, the number of edges with $\elabel$ that connect to attribute set $A$ is $|\edges(A, \elabel)| = |\{(v,\elabel_e,v')| (v,\elabel_e,v') \in \edges \land A(v') \supseteq A \land \elabel_e = \elabel\}|$.
The maximum numbers of matched vertices are computed by the following lemmas:

\begin{lemma}\label{lemma:maxpath1}
Given length $n$, edge label $\elabel$, and attribute set $A$, the maximum number of matched vertices for path patterns of $n-1$ length concatenated with $\elabel$ and $A$ 
is $|\edges(A, \elabel)|\cdot {(d_{m})}^{n-1})$ where $n>0$ and $d_{m}$ indicates the maximum in-degrees among vertices.
\end{lemma}
{\it Proof}: $|\edges(A_n, \elabel_{n})|$ indicates the number of edges with $\elabel_n$ connecting to vertices whose  attribute set includes $A_n$. Given $\langle A, \elabel_n, A_n \rangle$ as path pattern, the maximum number of vertices that match the path pattern is $|\edges(A_n, \elabel_{n})|$.
When the length of path patterns, the number of paths whose targets' attributes include $A_n$ exponentially increase by $d_m$.
Therefore, the maximum number of matched vertices for path patterns of $n-1$ length concatenated with $(\elabel_{n}, A_n)$ is $|\edges(A_n, \elabel_{n})|\cdot {(d_{m})}^{n-1})$.
\hfill{} $\square$

\begin{lemma}\label{lemma:maxpath2}
Given attribute set $A$ and edge label $\elabel$, the maximum number of matched vertices for $p=\langle A, \elabel, \ldots \rangle$ is $|\vertices(A, \elabel)|$.
\end{lemma}

{\it Proof}: This is obvious because the maximum number of vertices that can be sources of paths that match $\langle A, \elabel, \ldots \rangle$ is the number of vertices that have edges with $\elabel$. 
\hfill{} $\square$

From these lemma, we prune the candidates of edge labels and attribute sets of targets.
First, if $|\edges(A, \elabel)|\cdot {d_{m}}^{n-1}>\minsup$, $(A, \elabel)$ is a candidate of $\langle \label_n , A_n \rangle$ for path patterns with length $n$.
Second, unless $\elabel$ satisfies $\min(|\vertices(A, \elabel)|,|\edges(A, \elabel)|\cdot {d_{m}}^{\minlen-1}) \leq \minsup$ for any $A$,
$\elabel$ is a candidate of $\elabel_0$ and $\elabel^*$ for path patterns. 
We denote sets of edge labels and attributes of targets that could be involved in frequent path patterns by $\elabels_F$ and $\attributes_T$, respectively. 
The suffix pruning can reduce the candidates of edge labels and attribute sets that are added to frequent path patterns.




\smallskip
\noindent
{\bf Enhanced candidate generation.}
From the two anti-monotonic properties and suffix pruning, we can remove the candidates of path patterns if (1) the prefix of path pattern is not frequent, (2) the dominating path pattern is not frequent, and (3) suffix is not involved in frequent path patterns.
In the enhanced candidate generation, we extend path patterns in two ways.
{\it Vertical}; it first starts path patterns with zero length (i.e., frequent attributes), and then extends patterns to \minlen{} length by adding frequent the suffix. 
{\it Horizontal}; it finds unit path patterns (i.e., whose all attribute sets include a single attribute) and then combines attribute sets in frequent path patterns.
The pruned patterns can be excluded from candidates of path patterns without sacrificing the accuracy.

Our algorithm is modified based on the enhanced candidate generation.
The frequent attribute set discovery step contributes to finding $\elabels_F$ and $\attributes_T$.
We compute $\elabels_F$ and $\attributes_T$ by finding $\attributes_T$ for all edge labels and then $\elabels_F$, finally refining $\attributes_T$ by using $\elabels_F$.

In the frequent simple path pattern discovery step, we enumerate unit path patterns with one length. 
We vertically extend $\pathpatterns_0$ with one length by adding $\elabels_F$ and $\attributes_F$.
The candidates of path patterns can be reduced by using the suffix pruning.
After checking the frequency of all unit path patterns with one length, we horizontally combine two frequent paths to obtain path patterns whose attributes are more than one. 
This horizontal extension drastically reduces the number of candidates because if either two paths are not frequent, the combined paths are not frequent based on the pruning strategy in Lemma~\ref{lemma:antimonotonicity_2}.
We repeat vertical and horizontal extensions until obtaining frequent path patterns with $k$ length.


In the frequent reachability path patterns, we also use Lemma~\ref{lemma:antimonotonicity_2}; finding frequent reachability path patterns whose attribute set include a single attribute and them combined the two paths to obtain complex path patterns.

In the rule discovery, we use Lemmas~\ref{lemma:antimonotonicity_1} and \ref{lemma:antimonotonicity_2}. 
We first search for frequent rules that are combined one length unit path patterns.
Then, we vertically and horizontally extend the path patterns in frequent rules.
In more concretely, if $p \Rightarrow p'$ is frequent, $p_v \Rightarrow p'$, $p_h \Rightarrow p'$, $p \Rightarrow p'_v$, and $p \Rightarrow p'_h$ are candidates, 
where $p_v$ and $p_h$ are path patterns extended from $p$ vertically and horizontally, respectively.



\subsection{Auxiliary data structure}
Data structures have important roles for efficient rule discovery. 
Our algorithm maintains data structure to store pairs of a matched path pattern and targets of corresponding paths for each vertex.
We can grow paths by using the data structure without searching for paths from scratch.
In addition, our algorithm maintains pairs of dominating and dominated path patterns.
This is effective to generate candidates of rules; if $p \Rightarrow p'$ is frequent, we can obtain $p_v$, $p_h$, $p'_v$, and $p'_h$ by accessing the data structure.

\subsection{Pseudo-code and Complexity analysis}

Algorithm~\ref{alg:discovery} shows a pseudo code of our algorithm.
{\sf DiscoverX} functions search for vertices matched with candidates and find frequent patterns. 
Frequent attribute sets discovery step (lines 2--7) is generally the same as the basic Appriori algorithm and additionally computes frequent edges and attributes of targets.
Frequent simple path pattern discovery step (lines 8--12) vertically and horizontally extends the candidate of path patterns from already found frequent path patterns.
Frequent reachability path pattern discovery step (lines 13--17) conducts breadth-first search for each frequent edge label from each vertex and obtains the set of reached vertices $\vertices_T$. Then, it discovers frequent reachability path patterns.
It iteratively generates candidates of paths by horizontally extending the frequent reachability path patterns.
Rule discovery step (lines 18--22) generates the first candidates of rules which are any pairs of unit path patterns of one length simple and reachability path patterns. Then, it extends path patterns in the found rules horizontally and vertically.
Finally, it compute the metrics of frequent rules.

\begin{algorithm}[!t] 
	\caption{Algorithm for path association rule mining}	\label{alg:discovery}
		\DontPrintSemicolon
			    \SetKwInOut{Input}{input}
	            \SetKwInOut{Output}{output}
	             \SetKwFunction{Expand}{Expand}
	            \Input{Graph $\graph$, minimum support $\minsup$, maximum length $\minlen$}
	            \Output{set of rules \rules}
	            {\bf If} minimum support is relative {\bf Then} $\minsup \gets \minsup |\vertices|$\\
              ${\pathpatterns}_0,\ldots, {\pathpatterns}_k, {\pathpatterns}^*, \rules \gets \varnothing$\\
              
              ${\mathcal C^A} \gets \forall a \in \attributes$\\

              \While{${\mathcal C}^A \neq \varnothing$}{
                  ${\pathpatterns}_0 \gets {\pathpatterns}_0 \cup$ {\sf DiscoverAtt(${\mathcal C}^A, \graph, \minsup$)}\\
                  ${\mathcal C}^A \gets$ {\sf GenerateCandidate($\pathpatterns_0$)}\\
              }
                $(\elabels_F,\attributes_T) \gets$ {\sf MatchedEdges($\pathpatterns_0, \graph$)}\\
               
            \For{$i = 1$ to $k$}{
                  ${\mathcal C}^S \gets ${\sf VerticalExtend(${\pathpatterns}_{i-1},\elabels_F,\attributes_T$)\\       
                  \While{${\mathcal C}^S \neq \varnothing$}{
                  ${\pathpatterns}_i \gets {\pathpatterns}_i \cup$ {\sf DiscoverSimple(${\mathcal C}^S, \graph, \minsup$)}\\
                  ${\mathcal C}^S \gets$ {\sf HorizontalExtend(${\pathpatterns}_i$)}\\
                    }
                }   
            }
                
            $\vertices_T \gets${\sf BFS($\elabels_F, \graph, \minlen$)}\\
            ${\mathcal C}^* \gets \forall p$ computed by $\elabels_F, \attributes_T, \pathpatterns_0$\\
              \While{${\mathcal C}^* \neq \varnothing$}{
                  ${\pathpatterns}^* \gets {\pathpatterns}^* \cup$ {\sf DiscoverReachable(${\mathcal C}^*, \graph, \minsup, \vertices_T$)}\\
                  ${\mathcal C}^* \gets$ {\sf HorizontalExtend(${\pathpatterns}^*$)}\\
                }
                
             ${\mathcal C}^R \gets \forall (p \Rightarrow p')$ such that $p,p'$ are unit path patterns $\in \pathpatterns_1 \cup \pathpatterns^*$\\
                \While{${\mathcal C}^R \neq \varnothing$}{
                  $\rules \gets \rules \cup$ {\sf DiscoverRule(${\mathcal C}^R, \graph, \minsup$)}\\
                  ${\mathcal C}^R \gets \forall (p \Rightarrow p')$ so that either $p$ or $p'$ are horizontally and vertically extended path patterns in $\rules$.\\
              }
              {\sf ComputeMetrics}$(\rules)$\\

              {\bf return} the set of rules;\\
\end{algorithm}

We describe the time complexity in our algorithm.
We denote the numbers of candidates of attribute sets, simple path patterns, reachability path patterns, and rules by $\candidates^A_i$, $\candidates^S_i$, $\candidates^*_i$, and $\candidates^R_i$ respectively, where $i$ is the $i$-th loop in while process. $|\attributes_{F}|$ is the number of frequent attributes.

Frequent attribute discovery step takes the same time complexity of Apriori algorithm with finding frequent edges and targets.
This step takes $O(|\attributes| +|\edges||\vertices|+\sum_{i=2}|\vertices||\candidates^A_i|)$.
Next, the time complexity of frequent simple path pattern discovery depends on the number of frequent attributes and edge labels.
The time complexity is $O(|\attributes_{F}| |\elabels_{F}| |\attributes_{T}|+\sum_{i=2}|\vertices| |\candidates^S_i|)$.
The frequent reachability path discovery step takes similar complexity of previous step while it needs bread-first search.
It take $O(|\elabels_F|(|\vertices|+|\edges|)+|\attributes_F| |\elabels_F| |\attributes_T|+\sum_{i=2}|\vertices| |\candidates^*_i|)$.
The rule discovery step is combined two path patterns and extends patterns in found rules. 
It takes $O(|\vertices| |\pathpatterns_1||\pathpatterns^*|+ \sum_{i=2}|\vertices||\candidates^R_i|)$.

In total, the time complexity of our algorithm takes $O(|\vertices||\edges|+|\elabels_F|(|\vertices|+|\edges|+|\attributes_{F}| |\attributes_{T}|)+\sum_{i=2}|\vertices|(|\candidates^A_i|+|\candidates^S_i|+|\candidates^*_i|+|\candidates^R_i|))$.
This analysis indicates that the complexity of our algorithm highly depends on the sizes of candidates.

The space complexity of our algorithm is $O(|\vertices|+|\edges|+|\attributes| +|\candidates^A|+|\candidates^S|+|\candidates^*|+|\candidates^R|+\sum_i^k|\pathpatterns_i|+|\pathpatterns^*|+|\rules|)$, where $|\candidates^A|$, $|\candidates^S|$ $|\candidates^*|$, and $|\candidates^R|$ are the maximum sizes of $|\candidates^A_i|$, $|\candidates^S_i|$ $|\candidates^*_i|$, and $|\candidates^R_i|$ for all $i$, respectively.
Generally, the memory space does not become an issue empirically compared with the running time.





\section{Approximation}

We present two  approximation methods; candidate reduction and sampling.

\noindent
{\bf Candidate reduction}.
Our optimization strategy can reduce the candidates by the suffix pruning in Lemma~\ref{lemma:maxpath1}.
This suffix pruning safely removes the suffix without sacrificing the accuracy, but the suffix are not often involved in frequent path patterns. 
For scalable mining, we further reduce the candidates of the suffix that are not likely necessary to find frequent patterns.

\noindent
\underline{Methodology}.
We approximate the candidates of suffix based on Lemma~\ref{lemma:maxpath1}. 
Given candidate reduction factor $\psi$ ($0.0 \leq \psi \leq 1.0$), 
the suffix $\langle \elabel , A \rangle$ is removed from the candidates if 
$|\elabels(A, \elabel)| \cdot {(d_{m})}^{\psi(n-1)}$ is not larger than $\minsup$.

Lemma~\ref{lemma:maxpath1} uses the maximum degree to safely prune the candidates of suffixes that are not involved frequent paths. 
It often works well with small accuracy loss because most of vertices have much smaller degrees than the maximum degree.

\noindent
\underline{Theoretical analysis}.
The candidate reduction enables to reduce the candidate of suffix, but it causes missing results.
We discuss the probabilities that frequent path patterns are missing.

\begin{theorem}
Given frequent path pattern $p$, the missing probability of $p$ is 
$P\left( \minsup \geq \frac{|\vertices(p)|(d_{m})^{\psi(n-1)}}{p_m(d_p)^{n-1}}{}\right)$, where $p_m$ is a probability that vertex matches with $p$ among paths that match path patterns with suffix and $d_p$ is the average degree of actual paths.
\end{theorem}
{\it Proof}: We remove the suffix if $|\edges(A,\elabel)|{(d_{m})^{\psi(n-1)}}$ is not larger than $\minsup$.
We define $|P|$ as the number of paths from any sources to the target of suffix, and $|P| = \vertices(p)\cdot p_m$. Furthermore, $|P| = |\edges(A,\elabel)|(d_p)^{n-1}$.
We can derive the following probability:

\begin{eqnarray}
&P\left( \minsup \geq |\edges(A,\elabel)|{(d_{m})^{\psi(n-1)}}\right)\\
=& P\left( \minsup \geq \frac{|P|}{(d_p)^{n-1}}{(d_{m})^{\psi(n-1)}}\right)\\
=& P\left( \minsup \geq \frac{|\vertices(p)|(d_{m})^{\psi(n-1)}}{p_m(d_p)^{n-1}}{}\right)
\end{eqnarray}
\hfill{} $\square$

This theorem indicates that the missing probability decreases if (1) $|\vertices(p)|$ is large and (2) the difference between the maximum in-degree and actual average in-degree is large. 
In this approximation, we can reduce the number of candidates without false positives (i.e., guarantee precision is 1.0). 

\noindent
{\bf Sampling}.
Sampling is effective to reduce the computation cost in many data mining studies~\cite{fan2022discovering,lee1998sampling,toivonen1996sampling,fan2022parallel}.
In our sampling method, we pick a set of vertices to reduce the computation cost for finding matched vertices.

\noindent
\underline{Methodology}.
We need a sampling strategy that keeps the accuracy as much as possible. 
In our sampling strategy, we use a {\it stratified} sampling according to attributes of vertices~\cite{thompson2002sampling}.
We group the set of vertices into {\it strates} according to their attribute sets.
We remove vertices that have no frequent attributes, because they do not contribute to frequent path patterns.
From each strata, we pick vertices with $\rho$ sampling ratio.
We estimate the frequency of $p$ by the following:
\begin{equation}
     \widetilde{|\vertices(p)|} = \frac{|\vertices_s(p)|}{\rho}.
\end{equation}
where $\vertices_s$ denotes the set of sampled vertices in strata that is related to $A_0$ of $p$.

\noindent
\underline{Theoretical analysis}. The accuracy of sampling is represented by variance.
The variance of our sampling is as follows:
\begin{equation}
    s^2=\frac{\sum_{v_i\in \vertices_s} (x_i -\overline{x})}{\rho|\vertices_s|-1}.
\end{equation}
where $x_i$ is one if $v_i$ matches $p$, otherwise zero. $\overline{x}$ is $\frac{\widetilde{|\vertices_s(p)|}}{|\vertices_s|}$.

The confidence interval is the following:
\begin{equation}
     |\vertices_s|(\overline{x} - z \cdot \frac{s}{\sqrt{\rho|\vertices_s|}}) < |\vertices(p)| < |\vertices_s|(\overline{x} + z \cdot \frac{s}{\sqrt{\rho|\vertices_s|}})
\end{equation}
where $z$ indicates z-values for confidence intervals.

Compared with candidate reduction, the sampling possibly causes false positives, that is, infrequent path patterns could be included as frequent path patterns.
While, it can reduce the computation costs in a case that the candidate reduction does not work well. 


\section{Parallelization}
\label{sec:parallelization}

We accelerate our algorithm by parallelization.    
In our parallelization strategy, vertices are partitioned to balance the computing costs of threads.
Each vertex is associated with attributes and connects edges, each of them are either frequent or non-frequent.
Therefore, if some threads are assigned many frequent vertices, the threads could have large burdens.
We partition the set of vertices to balance their burdens.

Our approach is a cost-based partitioning.
Given the number of threads $N$, it estimates computing costs on vertices and then the set of vertices are divided into $N$ subsets so that the sum of costs in subsets are almost the same.
We describe how to estimate the cost and then partition the set of vertices.

\smallskip
\noindent
{\bf Cost estimation.}
The cost to find matched vertices increases the number of matched paths.
Thus, the cost is equal to the number of matched paths from each vertex.
First, we eliminate the unnecessary vertices that do not have frequent attributes and/or outgoing-edges with frequent edge labels.
These vertices obviously have no matched frequent path patterns. 
In other vertices, the number of matched paths depends on their attributes and out-going edges.
If their attributes and out-going edges are very frequent, vertices could have a large number of matched path patterns. 
We simply estimate the cost of vertex $C(v)$ as follows:
\begin{equation}
    C(v) = d_F(v)\cdot |A_F(v)|
\end{equation}
where $d_F(v)$ and $|A_F(v)|$ indicate the out-degree of edges with frequent edge labels and the number of frequent attributes on vertex $v$, respectively.
This estimation can be computed in $O(1)$ because $v$ does not have a large numbers of edges and attributes.

\smallskip
\noindent
{\bf Assignment.}
We divide the set of vertices into $N$ subsets according to the estimated costs.
To assign vertices, we use a greedy algorithm; estimates costs of vertices, sorts the vertices in ascending order of their costs, and repeatedly assigns the vertex with the largest cost among not-assigned ones to threads with the smallest sum of costs of assigned vertices, while we do not assign vertices to threads if threads are already assigned $|\vertices|/N$.
The time complexity of the assignment is $O(n |\vertices|\log |\vertices|)$.






\section{Experimental Study}
\label{sec:experiment}

We designed our experimental studies to show
(1) efficiency of our algorithm and (2) effectiveness of path association rule mining.

We implemented all algorithms by C++.
All experiments were performed on a Linux server with 512GB of memory and an Intel(R) Xeon(R) CPU E5-2699v3 processor. 

\noindent
{\bf Experimental Setting}.
We provide an overview of our experimental setup, including
datasets, compared methods, and parameters.


\begin{figure*}[!t]
  \subfloat[Nell]{\epsfig{file=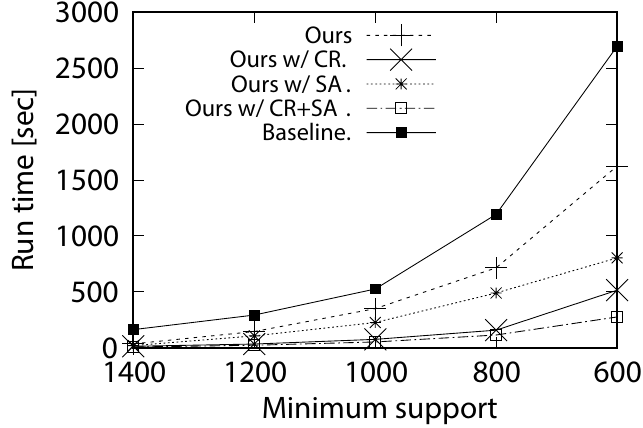,width=0.33\linewidth}
  }
  \subfloat[DBpedia]{\epsfig{file=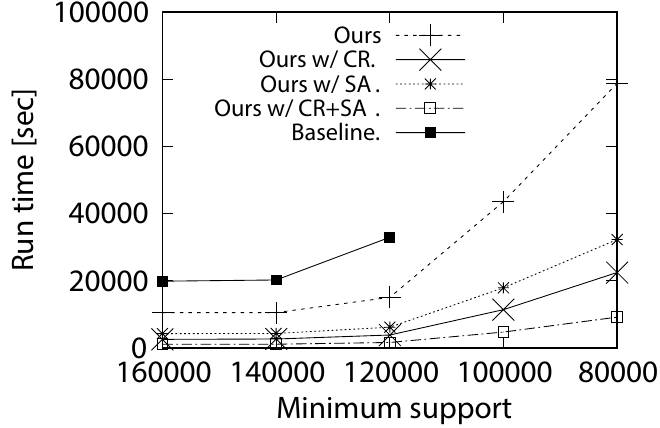,width=0.33\linewidth}
  }
  \subfloat[Pokec]{\epsfig{file=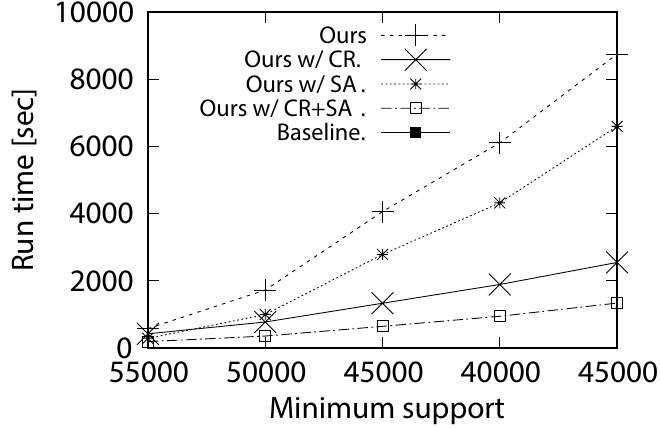,width=0.33\linewidth}
  }
   \vspace{-3mm}
    \caption{Impact of minimum support $\minsup$ to run time. The missing plots indicate that the methods did not finish within 24 hours.}
    \label{fig:runtime_minsupport}
    
\end{figure*}

\begin{figure*}[!t]
\vspace{-5mm}
  \subfloat[Nell]{\epsfig{file=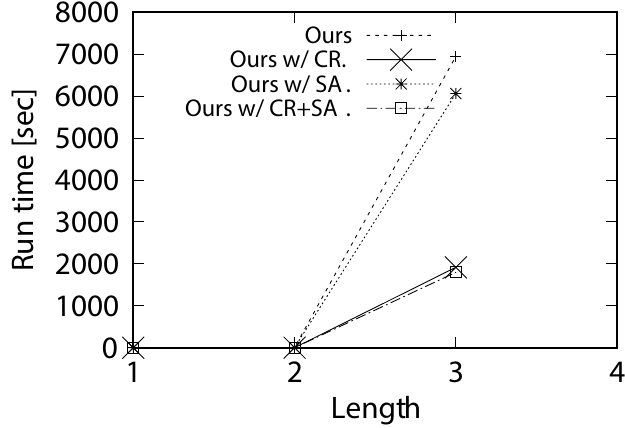,width=0.33\linewidth}
  }
  \subfloat[DBpedia]{\epsfig{file=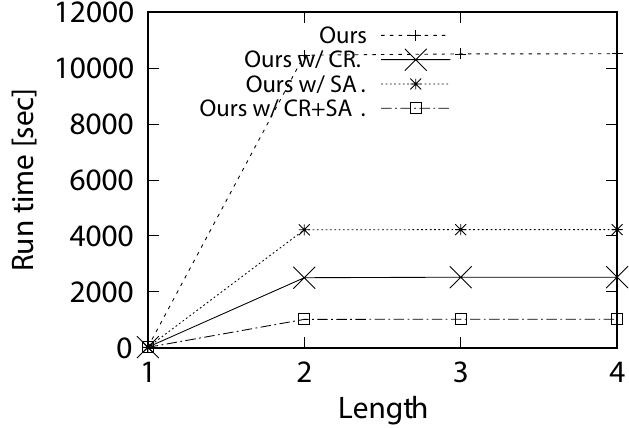,width=0.33\linewidth}
  }
  \subfloat[Pokec]{\epsfig{file=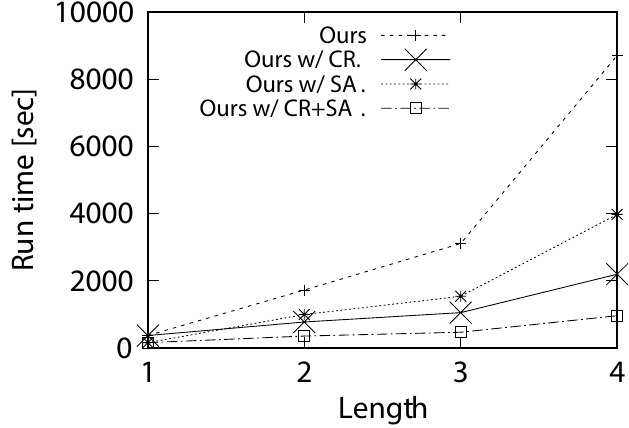,width=0.33\linewidth}
  }
   \vspace{-3mm}
    \caption{Impact of length to run time}
    \label{fig:runtime_length}
\end{figure*}

\begin{figure*}[!t]
\vspace{-5mm}
  \subfloat[Nell]{\epsfig{file=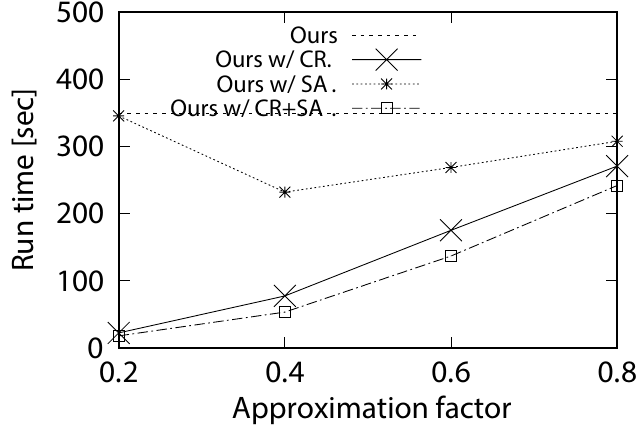,width=0.33\linewidth}
  }
  \subfloat[DBpedia]{\epsfig{file=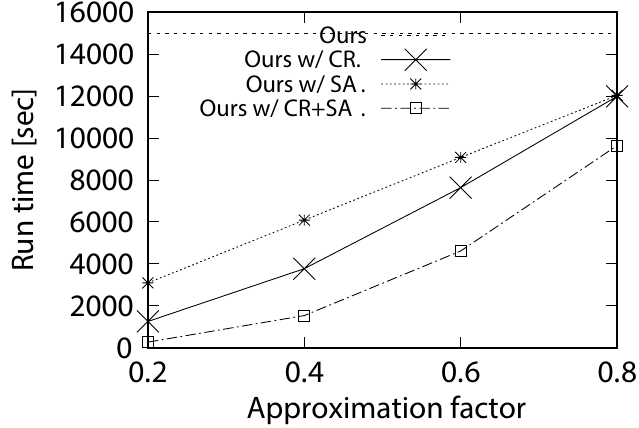,width=0.33\linewidth}
  }
  \subfloat[Pokec]{\epsfig{file=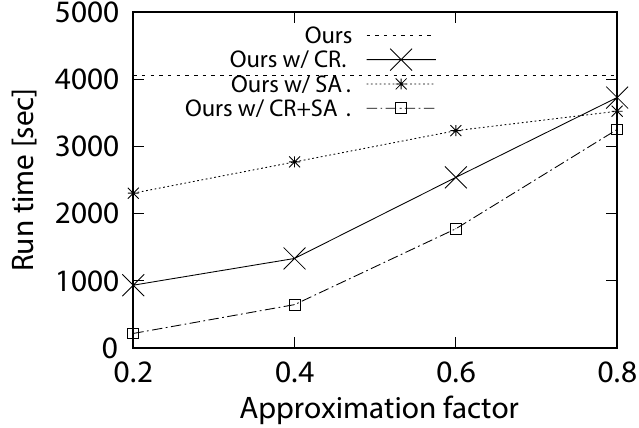,width=0.33\linewidth}
  }
   \vspace{-3mm}
    \caption{Impact of approximation factors to run time}
    \label{fig:runtime_approxiamtiohn}
\end{figure*}

\begin{figure}[!t]
    \begin{minipage}[t]{1.0\linewidth}
    \includegraphics[width=1.0\linewidth]{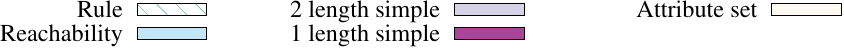}
    \end{minipage}
    \begin{minipage}[t]{1.0\linewidth}
    \center
    \includegraphics[width=0.7\linewidth]{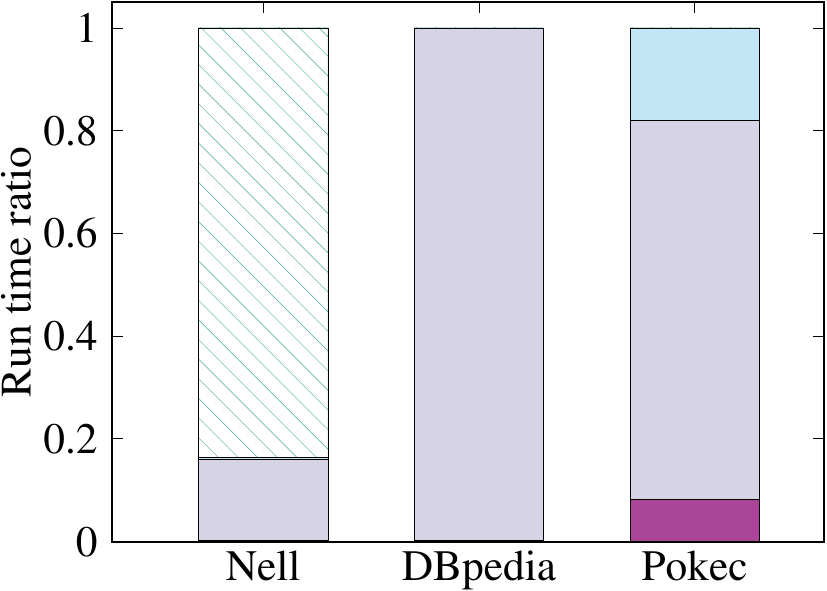}
    \caption{Run time ratio}
    \label{fig:runtime_ratio}
    \end{minipage}
\end{figure}

\noindent
\underline{\it{Dataset}}.
We use three datasets: Nell\footnote{\url{https://github.com/GemsLab/KGist}}, DBpedia$^1$, 
and Pokec\footnote{\url{https://snap.stanford.edu/data/soc-pokec.html}}.
\begin{itemize}
\item Nell~\cite{carlson2010toward} is a knowledge graph crawled from the web.
Its attributes are, for example, ceo, musician, company, and university, and edge labels are, for example, companyceo, competeswith, and workers.
\item DBpedia~\cite{auer2007dbpedia} is a knowledge graph extracted from wikipedia data.
Its attributes are, for example, actor, award, person, and place, and edge labels are, for example, child, spouse, and deathPlace.
\item Pokec is a social network service in which most users are Slovenian.
Its attributes are, for example, age, gender, city, and music, edge labels are, for example, follows, likes, and locatedIn. The number of edge label follows is very large, so we divide them according to out-degrees, such as small follows and large follows.
\end{itemize}
Table~\ref{table:datasets} shows data statistics.

\begin{table}[t]
\caption{Data statistics.}\label{table:datasets}
  \vspace{-3mm}
\scalebox{0.85}{
\begin{tabular}{lccccc}
\hline
          & $|\vertices|$ & $|\edges|$ & $|\elabels|$ & $|\attributes|$ & Avg. Attr.\\ \hline
Nell      & 46,682      &   231,634     & 821       & 266       & 1.5       \\
DBpedia   & 1,477,796   &   2,920,168   & 504       & 239       & 2.7       \\ 
Pokec     & 1,666,426   &  34,817,514 & 9            & 36302   & 1.1      \\ 
\hline
\end{tabular}
}
\end{table}

\noindent
\underline{\it{Parameters.}}
We set 2 as maximum path length and use 32 threads.
In minimum supports, we set absolute supports to each dataset; 1,000, 120,000, and 45,000 for Nell, DBpedia, and Pokec, respectively.
In approximation, we set 0.4 to candidate reduction factor $\psi$ and sampling rate $\rho$.
We vary these parameters to evaluate their impacts.
We compute absolute support, relative support, confidence, and lift.

\noindent
\underline{\it{Compared methods}}.
We compare variants of our algorithms because no methods cannot apply our problem. 
{\bf Baseline} is an algorithm that does not use the optimization strategy. 
{\bf Ours} is our algorithm without approximation methods. 
{\bf Ours w/ CR+SA} is our algorithm with both approximation methods, and {\bf Ours w/ CR} and {\bf Ours w/ SA} are our algorithms using either candidate reduction or sampling. 
All algorithms are parallelized.
Codes for graph pattern association rules on a single graph are not available~\cite{fan2017big,fan2015association,wang2020extending,fan2022discovering,fan2016adding}. 

\smallskip
\noindent
{\bf EXP-1 Efficiency}.
We show the efficiency of our algorithm.


\begin{table}[t]
\caption{Numbers of path patterns and rules.}\label{table:numberpatterns}
  \vspace{-3mm}
  \scalebox{0.85}{
\begin{tabular}{l|lll}
\hline
            & Nell & DBpedia & Pokec \\ \hline
Attribute set & 21 & 14 & 8\\
Simple path & 2283 & 6  & 10 \\
Reachability path & 35 & 6 & 31 \\
Rule       & 1,373,514 & 35 & 61 \\
\hline
\end{tabular}
}
\end{table}

\noindent
\underline{\it{Varying the minimum support $\minsup$}}.
Figure~\ref{fig:runtime_minsupport} shows run time of our algorithms by varying the minimum support in each dataset.
Our algorithms accelerate to find association rules compared with the baseline.
For example, in Pokec, Baseline did no finish within 24 hours because Pokec has a large number of attributes.
This indicates that our optimization strategy works well to reduce the candidates of path patterns.
In addition, approximation methods reduce computation costs largely.


\noindent
\underline{\it{Varying the number of threads}}.
Table~\ref{table:numberthreads} shows the run time by varying the number of threads in DBpedia and Pokec.
Our algorithm almost lineally decreases the run time as increasing threads.


\begin{table}[t]
\caption{Impact of \# of threads to run time [sec]}\label{table:numberthreads}
   \vspace{-3mm}
\scalebox{0.85}{
\begin{tabular}{l|lll|lll}
\hline
& \multicolumn{3}{|c|}{DBpedia} & \multicolumn{3}{|c}{Pokec}\\ 
             & 8     & 16 & 32 & 8     & 16 & 32\\ \hline
Ours  & 59,603    & 29,971 & 15,028 & 10,039     & 6,197 & 4,038\\
Ours w/ CR+SA   & 6,037  & 3,042 & 1,529 & 1,981 & 1,103 & 634\\
\hline
\end{tabular}
}
   \vspace{-3mm}
\end{table}






\begin{table}[t]
\caption{Accuracy.}\label{table:accuracy}
   \vspace{-3mm}
\scalebox{0.85}{
\begin{tabular}{l|ll|ll|ll}
\hline
& \multicolumn{2}{|c|}{Nell}& \multicolumn{2}{|c}{DBpedia} & \multicolumn{2}{|c}{Pokec}\\ 
            & Recall    & Precision & Recall     & Precision & Recall     & Precision\\ \hline
CR   & 0.233       & 1.0 & 1.0       & 1.0 & 1.0       & 1.0\\
SA      & 0.9999       & 0.999 & 1.0       & 1.0 &0.967&	0.983\\
CR+SA        & 0.233 &0.998 & 1.0       & 1.0 & 0.967&	0.983\\
\hline
\end{tabular}
}
\vspace{-3mm}
\end{table}



\noindent
\underline{\it{Varying the length $k$ of paths}} and approximation factors.

Figure~\ref{fig:runtime_length} shows the run time of each method varying the length $k$.
We set 1,800, 140,000, and 50,000 as $\minsup$.
We do not show the performance of the baseline because it does not finish within 24 hours when $k=3$ in all datasets.
Generally, as length increases, the run time increases because the numbers of rules and path patterns increase.
In Nell, the number of rules drastically increase when $k=4$, our algorithm did not finish within 24 hours.
In DBpedia, the number path patterns does not increase when $k$ is larger than 2, so the run time does not increase.
From these results, we can confirm our approximation methods work well when $k$ is large.

\noindent
\underline{\it{Varying the approximation factors}}.
Figure~\ref{fig:runtime_approxiamtiohn} shows the run time varying approximation factor.
The approximation factor indicates candidate reduction $\psi$ factor and sampling rate $\rho$ for CR and SA, respectively.
From this result, we can see that when the approximation factors are small, the run time is small.
However, in Nell, the run time of Our w/ SA is large when $\rho = 0.2$.
This is because the number of false negatives is large, so the frequent rule discovery step takes a large time.

\begin{table}[t]
\caption{Impact of approximation factors to accuracy in Nell}\label{table:accuracy_factor_nell}
   \vspace{-3mm}
\scalebox{0.85}{
\begin{tabular}{l|ll|ll|ll}
\hline
& \multicolumn{2}{|c|}{CR}& \multicolumn{2}{|c|}{SA} & \multicolumn{2}{|c}{CR+SA}\\ 
             & Recall     & Precision & Recall& Precision     & Recall & Precision\\ \hline
0.2	&0.0461	&1	&0.9999	&0.495	&0.046	&0.500\\
0.4	&0.233	&1	&0.9999	&0.999	&0.233	&0.998\\
0.6	&0.532	&1	&0.9999	&0.9999	&0.532	&0.9999\\
0.8	&0.797	&1	&0.9999	&0.9999	&0.797	&0.9999\\

\hline
\end{tabular}
}
\end{table}

\begin{table}[t]
\caption{Impact of approximation factors to accuracy in DBpedia}\label{table:accuracy_factor_dbpedia}
   \vspace{-3mm}
\scalebox{0.85}{
\begin{tabular}{l|ll|ll|ll}
\hline
& \multicolumn{2}{|c|}{CR}& \multicolumn{2}{|c|}{SA} & \multicolumn{2}{|c}{CR+SA}\\ 
             & Recall     & Precision & Recall& Precision     & Recall & Precision\\ \hline
0.2	&1.0	&1.0		&1.0	&1.0	&1.0	&1.0\\
0.4	&1.0	&1.0		&1.0	&1.0	&1.0	&1.0\\
0.6	&1.0	&1.0		&1.0	&1.0	&1.0	&1.0\\
0.8	&1.0	&1.0		&1.0	&1.0	&1.0	&1.0\\
\hline
\end{tabular}
}
\end{table}

\begin{table}[t]
\caption{Impact of approximation factors to accuracy in Pokec}\label{table:accuracy_factor_pokec}
   \vspace{-3mm}
\scalebox{0.85}{
\begin{tabular}{l|ll|ll|ll}
\hline
& \multicolumn{2}{|c|}{CR}& \multicolumn{2}{|c|}{SA} & \multicolumn{2}{|c}{CR+SA}\\ 
             & Recall     & Precision & Recall& Precision     & Recall & Precision\\ \hline
0.2	&1.0	&1.0	&0.934	&0.983	&0.934	&0.983\\
0.4	&1.0	&1.0	&0.967	&0.983	&0.967	&0.983\\
0.6	&1.0	&1.0	&0.984	&0.986	&0.984	&0.984\\
0.8	&1.0	&1.0	&0.984	&0.984	&0.984	&0.984\\
\hline
\end{tabular}
}
\end{table}

\noindent
{\bf EXP-2 Run time analysis}.
Table~\ref{fig:runtime_ratio} shows the ratio of run time on each step.
The ratio differs across datasets. Nell and DBpedia take their run time largely for finding rules and two length simple path patterns, respectively. In Pokec, it takes time for finding path patterns instead of rules.
Their run time ratios are according to the numbers of patterns and rules (see Table~\ref{table:numberpatterns}). These results are consistent with our complexity analysis, which the numbers of candidates highly affect to time complexity. 


\noindent
{\bf EXP-3 Accuracy}.
Table~\ref{table:accuracy} shows the recall and precision of our approximation methods compared to our exact algorithm (i.e., recall and precision are 1.0 if approximation methods returns the same results of exact algorithm).
In DBpedia and Pokec our approximation methods achieve very high accuracy, while in Nell the candidate reduction decreases the recall.
In Nell, the difference between the maximum and average in-degrees is small, so the missing probabilities become large. On the other hand, the differences in DBpedia and Pokec are large, so it effectively reduces the candidates.

Tables~\ref{table:accuracy_factor_nell}--\ref{table:accuracy_factor_pokec} show the recall and precision varying with approximation factors in each dataset.
We can see that the accuracy are quite high in DBpedia and pokec, while in Nell, CR and SA do not work well when approximation factors are small.
This is because the size of Nell is small compared with DBpedia and pokec, so the candidate reduction and sampling do not work well in some cases.

 \begin{figure*}[!t]
  \subfloat[Nell]{\epsfig{file=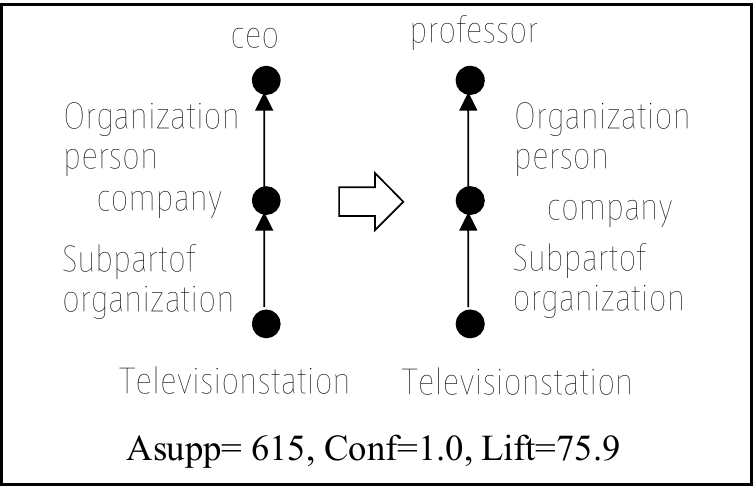,width=0.33\linewidth}
  }
  \subfloat[DBpedia]{\epsfig{file=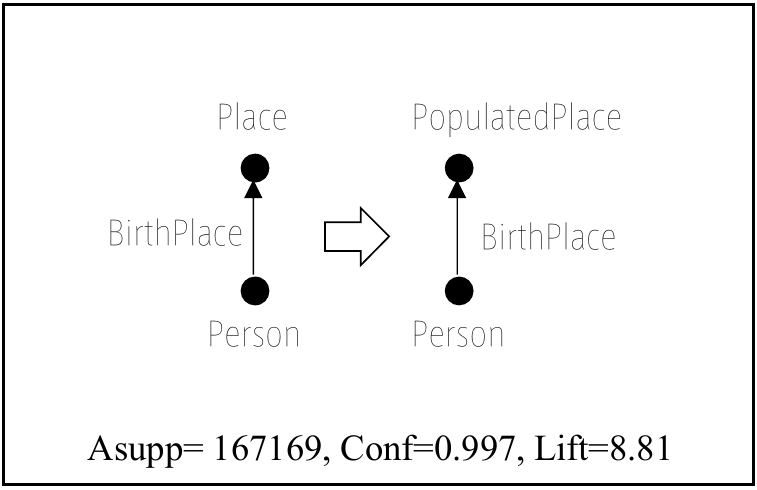,width=0.33\linewidth}
  }
\subfloat[Pokec]{\epsfig{file=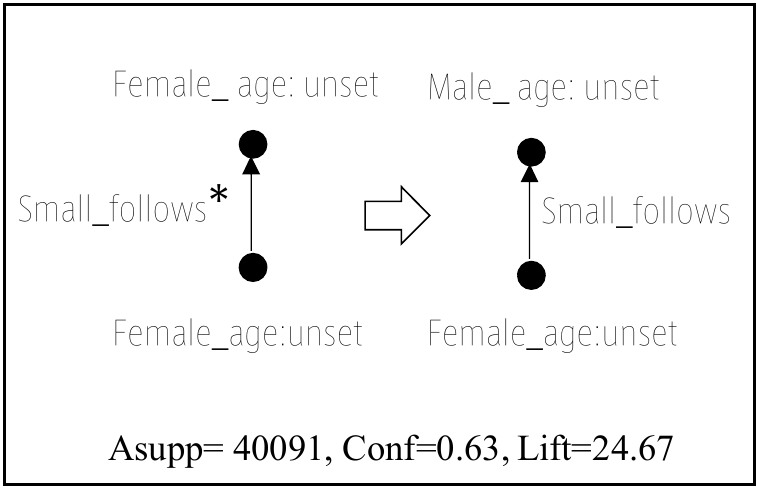,width=0.33\linewidth}
  }
  \vspace{-3mm}
    \caption{Examples of path association rules}
    \label{fig:experiment_examples}
    \vspace{-3mm}
\end{figure*}

\noindent
{\bf EXP-4 Effectiveness}.
We applied our method for knowledge extraction in three datasets and bias checking on Pokec.
Existing works cannot find these rules and compute metrics of them due to their semantics (e.g., we tried AnyBurl~\cite{meilicke2020reinforced}).
We note that our method does not have specific purposes such as knowledge graph completion, so we evaluate effectiveness qualitatively.

First, Figure~\ref{fig:experiment_examples} (a) illustrates a rule found in Nell;
This rule indicates ``television station'' is sub part of ``company'' that organized by ``CEO'' and ``professor''. 
Interestingly, there are no vertices associated with both CEO and professor on Nell, and ``company'' that organized by ``CEO'' and ``professor'' is not frequent.
This indicates ``company'' that have many ``television station'' is organized by at least different two people who are either CEO or professor.
This rule provides insights that such large companies hires professors.

Figure~\ref{fig:experiment_examples} (b) illustrates a rule found in DBpedia.
The confidence of this rule is quit high (i.e., 0.997).
On the other hand, the confidence of $\{\labelfont{Place} \} \Rightarrow \{\labelfont{PopulatedPlace} \}$ is 0.75.
This leads two possibilities of insights; (1) locations in which people recorded in DBpedia born become famous places and (2) knowledge extraction techniques in DBpedia register the locations as populated places because the name of locations often appear with the name of people in DBpedia.

In Pokec, we can find an interesting rule about reachability path patterns.
Figure~\ref{fig:experiment_examples} (c) illustrates the rule.
This rule shows the low probability that ``female who unset their ages connecting other female who unset their ages within two hops'' follow ``male who unset their ages''.
The confidences of other rules are more than 0.8, so this is the outlier case.
This rule helps to understand the friendship in social network service.

Second, we checked social bias in Pokec by our method.
We compare biases between females and males to educations of their friends. 
We compare $\langle \{\labelfont{gender} \} \rangle \Rightarrow \langle \{ \}, \labelfont{follows}, \{\labelfont{education:A} \} \rangle$, where $\labelfont{gender}$ and $A$ indicate male/female and the attributes related to educations, respectively.
We found 767 rules with $\minsup=500$.
We compare the support and confidence between these rules.
The numbers of vertices with male and female are 804{,}327 and 828{,}275, respectively, so the number of females is larger than that of males.
The sums of absolute supports on male and female are 464{,}725 and 466{,}725, and the sums of confidences are 2.445 and 2.064.
These results indicate that males more likely have friends who registered their educations than females, so there are gender biases in Pokec.
Our path association rules can be used to evaluate biases in property graphs.

\noindent
{\bf EXP-5 Ablation study}.
Table~\ref{table:ablation} shows the run time of our methods that we do not use some techniques.
``Ours w/o suffix pruning'' indicates our method that does not use suffix pruning.
``Ours w/o cost-based parallel'' indicates our method that randomly assigns the same numbers of vertices to threads without considering estimated costs.
From these results, Ours is drastically fast compared with Baseline.
The effectiveness of each technique depends on dataset.
For example, suffix pruning works well in Pokec, while does not in Nell. When the number of attributes is large, the suffix pruning works well.
In parallelization, in Nell and DBpedia, cost-based sampling slightly increase the efficiency, while in Pokec, slightly decreases. In Pokec, the costs of vertices are very similar, so the costs of threads become similar even if we do not estimate their costs.

\begin{table}
\caption{Ablation study}\label{table:ablation}
  \vspace{-3mm}
\scalebox{0.85}{
\begin{tabular}{l|l|l|l}
\hline
	&Nell	&DBpedia	&Pokec\\\hline
Ours	&341.7	&15028.3	&4038.5\\
Baseline	&525.4	&32907	& ---\\
Ours w/o suffix pruning	&304.4	&15466.4	&6171.3\\
Ours w/o cost-based parallel	&360.4	&15446.8	&4006.9\\
\hline
\end{tabular}
}
\end{table}

\section{Related Work}
\label{sec:related}
We review support measure on graphs and association rules. 
We explain additional studies in the supplementary file, such as graph pattern mining and matching and path query.


\noindent
{\bf Support measure on graphs.}
Support measures in items do not hold anti-monotonic property on a single graph.
Intuitively, the number of paths is often larger than the number of vertices though paths are more complex than vertices. 
To hold anti-monotonic properties, several support measures have been proposed; maximum independent set based support (MIS)~\cite{vanetik2002computing} minimum-image-based support (MNI)~\cite{bringmann2008frequent}, minimum clique partition (MCP)~\cite{calders2008anti}, minimum vertex cover (MVC)~\cite{meng2017flexible}, and maximum independent edge set support (MIES)~\cite{meng2017flexible}.
Their common semantic is to avoid duplicate counts of graph patterns if vertices are involved in multiple matches.
These support measures have three drawbacks.
First, they cannot be applied to relative support because it is hard to count the maximum numbers of graph patterns that possibly appear in a given graph.
Second, the time complexity is very large. For example, the problems of computing MIS and MNI are NP-hard.
Third, they are not intuitive because it is difficult to understand why vertices match graph patterns and the others do not.

As far as we know, support measure proposed in \cite{fan2015association} can be applied to relative support for a single large graph. They proposed vertex-centric support measure that counts the number of vertices that match a pivot in a subgraph.
The possible maximum absolute support is the number of vertices.
It is also intuitive, but it is inefficient because it needs isomorphic subgraph matching. 

\noindent
{\bf Graph pattern mining and matching.}
A number of algorithms have been developed for graph pattern matching~
\cite{elseidy2014grami, shelokar2014three}. 
Graph pattern mining algorithms possibly accelerate discovering graph association rules.
Similarly, algorithms for isomorphic subgraph matching, e.g., \cite{han2019efficient,fan2020extending},
are developed for efficiently discovering matched patterns in a single large graph.
These methods are not suitable for discovering frequent path patterns because they require to find subgraphs from scratch repeatedly.
Our algorithm maintains the targets of paths to efficiently extend the path patterns and uses anti-monotonic properties to reduce the number of candidates.
In addition, graph pattern mining and matching techniques do not handle reachability path patterns.

\noindent
{\bf Graph association rule mining. }
Graph association rule mining is divided into two categories; transactional graph data (i.e., a set of graphs) and a single large graph.
Algorithms for transactional graph data aim to find rules that are included in more than \minsup{} transactional graph data  (e.g., \cite{inokuchi2000apriori,ke2009efficient,yan2002gspan}).
On the other hand, algorithms for a single large graph aim to find rules that appear in a single graph more than \minsup~\cite{fan2017big,fan2015association,wang2020extending,fan2022discovering,fan2016adding}.
We note that methods for transactional graph data and single large graph are not interchangeable.



There are a few association rule mining for a large single graph~\cite{fan2017big,fan2015association,wang2020extending,fan2022discovering,fan2016adding}. To the best of our knowledge, no methods focus on paths and handle reachability patterns.
Association rules for graph patterns on a large single graph, called GPARs, were introduced in \cite{fan2015association}.
Their graph patterns use vertex-centric subgraphs.
Their association rules focus on specific graph patterns (1) consequent is a single edge and (2) the set of vertices in consequent is a subset of vertices in antecedent. 
It evaluates whether an edge type specified by consequent is included in antecedent or not. 
In addition, their algorithm aims to find diversified association rules with a fixed consequent, instead of finding all frequent rules.

The following works extended or generalized the original GPAR.
Wang et al. \cite{wang2020extending} used different semantics from the original GPAR \cite{fan2015association}; (1) support measure is MSI, (2) both antecedent and consequent are subgraphs that include at least one edge, and (3) antecedent and consequent have no common edges. 
Their algorithm finds all frequent association rules.
The drawbacks of their rules are that (1) it cannot use relative support due to MSI and (2) it cannot find regularities between graph patterns and properties of vertices (e.g., occupation and gender) because both antecedent and consequent must have at least one edge and no common edges (i.e., we cannot specify a single property as consequent).
Fan et al. \cite{fan2016adding} extended the original GPAR to find quantified graph association rules, called QGAR, which handles potential edges and quantities of edges in graphs. 
QGAR can be considered as a general version of GPAR in terms of graph patterns.
This work mainly focuses on how to find quantified subgraphs efficiently, though they did not propose a sophisticated algorithm for QGAR. They find association rules in a naive way after finding quantified subgraphs.  
Fan et al. \cite{fan2022discovering} proposed graph association rules, called GARs, which are generalized cases of GPARs with vertex attributes.
Their algorithm aims to find the top-$m$ GARs with the largest frequency. In addition, to accelerate run time, it reduces a given graph by sampling and machine learning based on specific targets of applications.
The difference between our and its sampling is that our sampling reduces the candidate of source vertices instead that they reduce the graph itself.
GAR allows a single edge or attribute in the consequent, so the flexibility of rules is small compared with rules of \cite{wang2020extending,fan2016adding}.
In \cite{namaki2017discovering}, GPAR was applied to temporal graphs for discovering temporal regularities on dynamic graphs.
We here note that only GAR~\cite{fan2022discovering} handle general property graphs.

\smallskip
\noindent
{\bf Related rule mining. }
Numerous studies extract rules from graphs, such as graph functional dependency~\cite{fan2020discovering} and Horn rules~\cite{manola2004rdf,schmitz2006mining,galarraga2013amie,chen2016scalekb,meilicke2020reinforced,ortona2018robust}.
In particular, Horn rules are similar to path association rules. Horn rules can be considered as rules that a consequent is a single edge and its vertices in the antecedent (i.e., the specialized GPAR). 
They are not applied to graph association rule mining in general property graphs.

Subsequence mining aims to find regularities between sub-sequence patterns from sequences  (e.g.,~\cite{agrawal1995mining,zaki1998planmine,nowozin2007discriminative,fournier2015mining,fournier2014erminer}). 
They can be considered as a special type of graph association rule mining because a sequence can be considered as a path graph.
Subsequence mining directly cannot apply graph association rule mining because they cannot handle complex graphs. 

\smallskip
\noindent
{\bf Summary.}
Our problem differs from the prior works as follows:
(1) The path association rule is flexible in terms of consequent.
(2) We first apply the reachability path patterns to graph association rule mining. 
(3) We study the discovery of association rules on general property graphs, so it differs from existing works for non-property graphs (e.g., \cite{fan2015association,fan2016adding,wang2020extending}) and RDF (e.g., \cite{galarraga2013amie,chen2016scalekb,meilicke2020reinforced,ortona2018robust}).

\section{Concluding remarks}
\label{sec:conclusion}

We proposed a new concept, path association rule mining, which aims to find regularities in paths in a single large graph.
We developed an efficient and scalable algorithm for path association rule mining.
In our experimental study, we validated that path association rule mining discovers interesting insights and our algorithm efficiently finds the rules compared with baselines.


We have several research direction.
The path association rule mining aims to find all frequent path patterns.
We will extend it to find top-k interesting rules with new metrics that can capture the characteristics of path patterns. 
We also plan to apply path association rule mining in real-world applications.


\bibliographystyle{ACM-Reference-Format}
\bibliography{graphassociationrule}


\begin{thebibliography}{49}


\ifx \showCODEN    \undefined \def \showCODEN     #1{\unskip}     \fi
\ifx \showDOI      \undefined \def \showDOI       #1{#1}\fi
\ifx \showISBNx    \undefined \def \showISBNx     #1{\unskip}     \fi
\ifx \showISBNxiii \undefined \def \showISBNxiii  #1{\unskip}     \fi
\ifx \showISSN     \undefined \def \showISSN      #1{\unskip}     \fi
\ifx \showLCCN     \undefined \def \showLCCN      #1{\unskip}     \fi
\ifx \shownote     \undefined \def \shownote      #1{#1}          \fi
\ifx \showarticletitle \undefined \def \showarticletitle #1{#1}   \fi
\ifx \showURL      \undefined \def \showURL       {\relax}        \fi
\providecommand\bibfield[2]{#2}
\providecommand\bibinfo[2]{#2}
\providecommand\natexlab[1]{#1}
\providecommand\showeprint[2][]{arXiv:#2}

\bibitem[\protect\citeauthoryear{Agrawal, Imieli{\'n}ski, and Swami}{Agrawal
  et~al\mbox{.}}{1993}]%
        {agrawal1993mining}
\bibfield{author}{\bibinfo{person}{Rakesh Agrawal}, \bibinfo{person}{Tomasz
  Imieli{\'n}ski}, {and} \bibinfo{person}{Arun Swami}.}
  \bibinfo{year}{1993}\natexlab{}.
\newblock \showarticletitle{Mining association rules between sets of items in
  large databases}. In \bibinfo{booktitle}{\emph{SIGMOD}}.
  \bibinfo{pages}{207--216}.
\newblock


\bibitem[\protect\citeauthoryear{Agrawal and Srikant}{Agrawal and
  Srikant}{1995}]%
        {agrawal1995mining}
\bibfield{author}{\bibinfo{person}{Rakesh Agrawal} {and}
  \bibinfo{person}{Ramakrishnan Srikant}.} \bibinfo{year}{1995}\natexlab{}.
\newblock \showarticletitle{Mining sequential patterns}. In
  \bibinfo{booktitle}{\emph{Proceedings of the eleventh international
  conference on data engineering}}. \bibinfo{pages}{3--14}.
\newblock


\bibitem[\protect\citeauthoryear{Auer, Bizer, Kobilarov, Lehmann, Cyganiak, and
  Ives}{Auer et~al\mbox{.}}{2007}]%
        {auer2007dbpedia}
\bibfield{author}{\bibinfo{person}{S{\"o}ren Auer}, \bibinfo{person}{Christian
  Bizer}, \bibinfo{person}{Georgi Kobilarov}, \bibinfo{person}{Jens Lehmann},
  \bibinfo{person}{Richard Cyganiak}, {and} \bibinfo{person}{Zachary Ives}.}
  \bibinfo{year}{2007}\natexlab{}.
\newblock \showarticletitle{Dbpedia: A nucleus for a web of open data}.
\newblock In \bibinfo{booktitle}{\emph{The semantic web}}.
  \bibinfo{pages}{722--735}.
\newblock


\bibitem[\protect\citeauthoryear{Bonifati, Martens, and Timm}{Bonifati
  et~al\mbox{.}}{2019}]%
        {BonifatiMT19}
\bibfield{author}{\bibinfo{person}{Angela Bonifati}, \bibinfo{person}{Wim
  Martens}, {and} \bibinfo{person}{Thomas Timm}.}
  \bibinfo{year}{2019}\natexlab{}.
\newblock \showarticletitle{Navigating the Maze of Wikidata Query Logs}. In
  \bibinfo{booktitle}{\emph{{WWW}}}. \bibinfo{pages}{127--138}.
\newblock


\bibitem[\protect\citeauthoryear{Bonifati, Martens, and Timm}{Bonifati
  et~al\mbox{.}}{2020}]%
        {bonifati2020analytical}
\bibfield{author}{\bibinfo{person}{Angela Bonifati}, \bibinfo{person}{Wim
  Martens}, {and} \bibinfo{person}{Thomas Timm}.}
  \bibinfo{year}{2020}\natexlab{}.
\newblock \showarticletitle{An analytical study of large {SPARQL} query logs}.
\newblock \bibinfo{journal}{\emph{The VLDB Journal}} (\bibinfo{year}{2020}),
  \bibinfo{pages}{655--679}.
\newblock


\bibitem[\protect\citeauthoryear{Bringmann and Nijssen}{Bringmann and
  Nijssen}{2008}]%
        {bringmann2008frequent}
\bibfield{author}{\bibinfo{person}{Bj{\"o}rn Bringmann} {and}
  \bibinfo{person}{Siegfried Nijssen}.} \bibinfo{year}{2008}\natexlab{}.
\newblock \showarticletitle{What is frequent in a single graph?}. In
  \bibinfo{booktitle}{\emph{PAKDD}}. \bibinfo{pages}{858--863}.
\newblock


\bibitem[\protect\citeauthoryear{Calders, Ramon, and Van~Dyck}{Calders
  et~al\mbox{.}}{2008}]%
        {calders2008anti}
\bibfield{author}{\bibinfo{person}{Toon Calders}, \bibinfo{person}{Jan Ramon},
  {and} \bibinfo{person}{Dries Van~Dyck}.} \bibinfo{year}{2008}\natexlab{}.
\newblock \showarticletitle{Anti-monotonic overlap-graph support measures}. In
  \bibinfo{booktitle}{\emph{ICDM}}. \bibinfo{pages}{73--82}.
\newblock


\bibitem[\protect\citeauthoryear{Carlson, Betteridge, Kisiel, Settles,
  Hruschka, and Mitchell}{Carlson et~al\mbox{.}}{2010}]%
        {carlson2010toward}
\bibfield{author}{\bibinfo{person}{Andrew Carlson}, \bibinfo{person}{Justin
  Betteridge}, \bibinfo{person}{Bryan Kisiel}, \bibinfo{person}{Burr Settles},
  \bibinfo{person}{Estevam~R Hruschka}, {and} \bibinfo{person}{Tom~M
  Mitchell}.} \bibinfo{year}{2010}\natexlab{}.
\newblock \showarticletitle{Toward an architecture for never-ending language
  learning}. In \bibinfo{booktitle}{\emph{AAAI}}. \bibinfo{pages}{1306--1313}.
\newblock


\bibitem[\protect\citeauthoryear{Chen, Wang, and Goldberg}{Chen
  et~al\mbox{.}}{2016}]%
        {chen2016scalekb}
\bibfield{author}{\bibinfo{person}{Yang Chen}, \bibinfo{person}{Daisy~Zhe
  Wang}, {and} \bibinfo{person}{Sean Goldberg}.}
  \bibinfo{year}{2016}\natexlab{}.
\newblock \showarticletitle{ScaLeKB: scalable learning and inference over large
  knowledge bases}.
\newblock \bibinfo{journal}{\emph{The VLDB Journal}} \bibinfo{volume}{25},
  \bibinfo{number}{6} (\bibinfo{year}{2016}), \bibinfo{pages}{893--918}.
\newblock


\bibitem[\protect\citeauthoryear{Dastin}{Dastin}{2018}]%
        {bias2018}
\bibfield{author}{\bibinfo{person}{Jeffrey Dastin}.}
  \bibinfo{year}{2018}\natexlab{}.
\newblock \showarticletitle{Rpt-insight-amazon scraps secret ai recruiting tool
  that showed bias against women}.
\newblock \bibinfo{journal}{\emph{Reuters}} (\bibinfo{year}{2018}).
\newblock


\bibitem[\protect\citeauthoryear{Davidson, Liebald, Liu, Nandy, Van~Vleet,
  Gargi, Gupta, He, Lambert, Livingston, et~al\mbox{.}}{Davidson
  et~al\mbox{.}}{2010}]%
        {davidson2010youtube}
\bibfield{author}{\bibinfo{person}{James Davidson}, \bibinfo{person}{Benjamin
  Liebald}, \bibinfo{person}{Junning Liu}, \bibinfo{person}{Palash Nandy},
  \bibinfo{person}{Taylor Van~Vleet}, \bibinfo{person}{Ullas Gargi},
  \bibinfo{person}{Sujoy Gupta}, \bibinfo{person}{Yu He}, \bibinfo{person}{Mike
  Lambert}, \bibinfo{person}{Blake Livingston}, {et~al\mbox{.}}}
  \bibinfo{year}{2010}\natexlab{}.
\newblock \showarticletitle{The YouTube video recommendation system}. In
  \bibinfo{booktitle}{\emph{RecSys}}. \bibinfo{pages}{293--296}.
\newblock


\bibitem[\protect\citeauthoryear{Elseidy, Abdelhamid, Skiadopoulos, and
  Kalnis}{Elseidy et~al\mbox{.}}{2014}]%
        {elseidy2014grami}
\bibfield{author}{\bibinfo{person}{Mohammed Elseidy}, \bibinfo{person}{Ehab
  Abdelhamid}, \bibinfo{person}{Spiros Skiadopoulos}, {and}
  \bibinfo{person}{Panos Kalnis}.} \bibinfo{year}{2014}\natexlab{}.
\newblock \showarticletitle{Grami: Frequent subgraph and pattern mining in a
  single large graph}.
\newblock \bibinfo{journal}{\emph{PVLDB}} \bibinfo{volume}{7},
  \bibinfo{number}{7} (\bibinfo{year}{2014}), \bibinfo{pages}{517--528}.
\newblock


\bibitem[\protect\citeauthoryear{Fan, Fan, Li, Lu, Tian, and Zhou}{Fan
  et~al\mbox{.}}{2020a}]%
        {fan2020extending}
\bibfield{author}{\bibinfo{person}{Grace Fan}, \bibinfo{person}{Wenfei Fan},
  \bibinfo{person}{Yuanhao Li}, \bibinfo{person}{Ping Lu},
  \bibinfo{person}{Chao Tian}, {and} \bibinfo{person}{Jingren Zhou}.}
  \bibinfo{year}{2020}\natexlab{a}.
\newblock \showarticletitle{Extending Graph Patterns with Conditions}. In
  \bibinfo{booktitle}{\emph{SIGMOD}}. \bibinfo{pages}{715--729}.
\newblock


\bibitem[\protect\citeauthoryear{Fan, Fu, Jin, Lu, and Tian}{Fan
  et~al\mbox{.}}{2022a}]%
        {fan2022discovering}
\bibfield{author}{\bibinfo{person}{Wenfei Fan}, \bibinfo{person}{Wenzhi Fu},
  \bibinfo{person}{Ruochun Jin}, \bibinfo{person}{Ping Lu}, {and}
  \bibinfo{person}{Chao Tian}.} \bibinfo{year}{2022}\natexlab{a}.
\newblock \showarticletitle{Discovering association rules from big graphs}.
\newblock \bibinfo{journal}{\emph{PVLDB}} \bibinfo{volume}{15},
  \bibinfo{number}{7} (\bibinfo{year}{2022}), \bibinfo{pages}{1479--1492}.
\newblock


\bibitem[\protect\citeauthoryear{Fan, Han, Wang, and Xie}{Fan
  et~al\mbox{.}}{2022b}]%
        {fan2022parallel}
\bibfield{author}{\bibinfo{person}{Wenfei Fan}, \bibinfo{person}{Ziyan Han},
  \bibinfo{person}{Yaoshu Wang}, {and} \bibinfo{person}{Min Xie}.}
  \bibinfo{year}{2022}\natexlab{b}.
\newblock \showarticletitle{Parallel Rule Discovery from Large Datasets by
  Sampling}. In \bibinfo{booktitle}{\emph{SIGMOD}}. \bibinfo{pages}{384--398}.
\newblock


\bibitem[\protect\citeauthoryear{Fan and Hu}{Fan and Hu}{2017}]%
        {fan2017big}
\bibfield{author}{\bibinfo{person}{Wenfei Fan} {and} \bibinfo{person}{Chunming
  Hu}.} \bibinfo{year}{2017}\natexlab{}.
\newblock \showarticletitle{Big graph analyses: From queries to dependencies
  and association rules}.
\newblock \bibinfo{journal}{\emph{Data Science and Engineering}}
  \bibinfo{volume}{2}, \bibinfo{number}{1} (\bibinfo{year}{2017}),
  \bibinfo{pages}{36--55}.
\newblock


\bibitem[\protect\citeauthoryear{Fan, Hu, Liu, and Lu}{Fan
  et~al\mbox{.}}{2020b}]%
        {fan2020discovering}
\bibfield{author}{\bibinfo{person}{Wenfei Fan}, \bibinfo{person}{Chunming Hu},
  \bibinfo{person}{Xueli Liu}, {and} \bibinfo{person}{Ping Lu}.}
  \bibinfo{year}{2020}\natexlab{b}.
\newblock \showarticletitle{Discovering graph functional dependencies}.
\newblock \bibinfo{journal}{\emph{TODS}} \bibinfo{volume}{45},
  \bibinfo{number}{3} (\bibinfo{year}{2020}), \bibinfo{pages}{1--42}.
\newblock


\bibitem[\protect\citeauthoryear{Fan, Wang, Wu, and Xu}{Fan
  et~al\mbox{.}}{2015}]%
        {fan2015association}
\bibfield{author}{\bibinfo{person}{Wenfei Fan}, \bibinfo{person}{Xin Wang},
  \bibinfo{person}{Yinghui Wu}, {and} \bibinfo{person}{Jingbo Xu}.}
  \bibinfo{year}{2015}\natexlab{}.
\newblock \showarticletitle{Association rules with graph patterns}.
\newblock \bibinfo{journal}{\emph{PVLDB}} \bibinfo{volume}{8},
  \bibinfo{number}{12} (\bibinfo{year}{2015}), \bibinfo{pages}{1502--1513}.
\newblock


\bibitem[\protect\citeauthoryear{Fan, Wu, and Xu}{Fan et~al\mbox{.}}{2016}]%
        {fan2016adding}
\bibfield{author}{\bibinfo{person}{Wenfei Fan}, \bibinfo{person}{Yinghui Wu},
  {and} \bibinfo{person}{Jingbo Xu}.} \bibinfo{year}{2016}\natexlab{}.
\newblock \showarticletitle{Adding counting quantifiers to graph patterns}. In
  \bibinfo{booktitle}{\emph{SIGMOD}}. \bibinfo{pages}{1215--1230}.
\newblock


\bibitem[\protect\citeauthoryear{Fournier-Viger, Gueniche, Zida, and
  Tseng}{Fournier-Viger et~al\mbox{.}}{2014}]%
        {fournier2014erminer}
\bibfield{author}{\bibinfo{person}{Philippe Fournier-Viger},
  \bibinfo{person}{Ted Gueniche}, \bibinfo{person}{Souleymane Zida}, {and}
  \bibinfo{person}{Vincent~S Tseng}.} \bibinfo{year}{2014}\natexlab{}.
\newblock \showarticletitle{ERMiner: sequential rule mining using equivalence
  classes}. In \bibinfo{booktitle}{\emph{International Symposium on Intelligent
  Data Analysis}}. \bibinfo{pages}{108--119}.
\newblock


\bibitem[\protect\citeauthoryear{Fournier-Viger, Wu, Tseng, Cao, and
  Nkambou}{Fournier-Viger et~al\mbox{.}}{2015}]%
        {fournier2015mining}
\bibfield{author}{\bibinfo{person}{Philippe Fournier-Viger},
  \bibinfo{person}{Cheng-Wei Wu}, \bibinfo{person}{Vincent~S Tseng},
  \bibinfo{person}{Longbing Cao}, {and} \bibinfo{person}{Roger Nkambou}.}
  \bibinfo{year}{2015}\natexlab{}.
\newblock \showarticletitle{Mining partially-ordered sequential rules common to
  multiple sequences}.
\newblock \bibinfo{journal}{\emph{TKDE}} \bibinfo{volume}{27},
  \bibinfo{number}{8} (\bibinfo{year}{2015}), \bibinfo{pages}{2203--2216}.
\newblock


\bibitem[\protect\citeauthoryear{Gal{\'a}rraga, Teflioudi, Hose, and
  Suchanek}{Gal{\'a}rraga et~al\mbox{.}}{2013}]%
        {galarraga2013amie}
\bibfield{author}{\bibinfo{person}{Luis~Antonio Gal{\'a}rraga},
  \bibinfo{person}{Christina Teflioudi}, \bibinfo{person}{Katja Hose}, {and}
  \bibinfo{person}{Fabian Suchanek}.} \bibinfo{year}{2013}\natexlab{}.
\newblock \showarticletitle{AMIE: association rule mining under incomplete
  evidence in ontological knowledge bases}. In \bibinfo{booktitle}{\emph{WWW}}.
  \bibinfo{pages}{413--422}.
\newblock


\bibitem[\protect\citeauthoryear{Haller and Hadler}{Haller and Hadler}{2006}]%
        {haller2006social}
\bibfield{author}{\bibinfo{person}{Max Haller} {and} \bibinfo{person}{Markus
  Hadler}.} \bibinfo{year}{2006}\natexlab{}.
\newblock \showarticletitle{How social relations and structures can produce
  happiness and unhappiness: An international comparative analysis}.
\newblock \bibinfo{journal}{\emph{Social indicators research}}
  \bibinfo{volume}{75}, \bibinfo{number}{2} (\bibinfo{year}{2006}),
  \bibinfo{pages}{169--216}.
\newblock


\bibitem[\protect\citeauthoryear{Han, Kim, Gu, Park, and Han}{Han
  et~al\mbox{.}}{2019}]%
        {han2019efficient}
\bibfield{author}{\bibinfo{person}{Myoungji Han}, \bibinfo{person}{Hyunjoon
  Kim}, \bibinfo{person}{Geonmo Gu}, \bibinfo{person}{Kunsoo Park}, {and}
  \bibinfo{person}{Wook-Shin Han}.} \bibinfo{year}{2019}\natexlab{}.
\newblock \showarticletitle{Efficient subgraph matching: Harmonizing dynamic
  programming, adaptive matching order, and failing set together}. In
  \bibinfo{booktitle}{\emph{SIGMOD}}. \bibinfo{pages}{1429--1446}.
\newblock


\bibitem[\protect\citeauthoryear{House, Landis, and Umberson}{House
  et~al\mbox{.}}{1988}]%
        {house1988social}
\bibfield{author}{\bibinfo{person}{James~S House}, \bibinfo{person}{Karl~R
  Landis}, {and} \bibinfo{person}{Debra Umberson}.}
  \bibinfo{year}{1988}\natexlab{}.
\newblock \showarticletitle{Social relationships and health}.
\newblock \bibinfo{journal}{\emph{Science}} \bibinfo{volume}{241},
  \bibinfo{number}{4865} (\bibinfo{year}{1988}), \bibinfo{pages}{540--545}.
\newblock


\bibitem[\protect\citeauthoryear{Inokuchi, Washio, and Motoda}{Inokuchi
  et~al\mbox{.}}{[n.d.]}]%
        {inokuchi2000apriori}
\bibfield{author}{\bibinfo{person}{Akihiro Inokuchi}, \bibinfo{person}{Takashi
  Washio}, {and} \bibinfo{person}{Hiroshi Motoda}.}
  \bibinfo{year}{[n.d.]}\natexlab{}.
\newblock \showarticletitle{An apriori-based algorithm for mining frequent
  substructures from graph data}. In \bibinfo{booktitle}{\emph{PKDD}}.
  \bibinfo{pages}{13--23}.
\newblock


\bibitem[\protect\citeauthoryear{Kaur and Kang}{Kaur and Kang}{2016}]%
        {kaur2016market}
\bibfield{author}{\bibinfo{person}{Manpreet Kaur} {and}
  \bibinfo{person}{Shivani Kang}.} \bibinfo{year}{2016}\natexlab{}.
\newblock \showarticletitle{Market Basket Analysis: Identify the changing
  trends of market data using association rule mining}.
\newblock \bibinfo{journal}{\emph{Procedia computer science}}
  \bibinfo{volume}{85} (\bibinfo{year}{2016}), \bibinfo{pages}{78--85}.
\newblock


\bibitem[\protect\citeauthoryear{Ke, Cheng, and Yu}{Ke et~al\mbox{.}}{2009}]%
        {ke2009efficient}
\bibfield{author}{\bibinfo{person}{Yiping Ke}, \bibinfo{person}{James Cheng},
  {and} \bibinfo{person}{Jeffrey~Xu Yu}.} \bibinfo{year}{2009}\natexlab{}.
\newblock \showarticletitle{Efficient discovery of frequent correlated subgraph
  pairs}. In \bibinfo{booktitle}{\emph{ICDM}}. \bibinfo{pages}{239--248}.
\newblock


\bibitem[\protect\citeauthoryear{Lee, Kim, and Rhee}{Lee et~al\mbox{.}}{2001}]%
        {lee2001web}
\bibfield{author}{\bibinfo{person}{C-H Lee}, \bibinfo{person}{Y-H Kim}, {and}
  \bibinfo{person}{P-K Rhee}.} \bibinfo{year}{2001}\natexlab{}.
\newblock \showarticletitle{Web personalization expert with combining
  collaborative filtering and association rule mining technique}.
\newblock \bibinfo{journal}{\emph{Expert Systems with Applications}}
  \bibinfo{volume}{21}, \bibinfo{number}{3} (\bibinfo{year}{2001}),
  \bibinfo{pages}{131--137}.
\newblock


\bibitem[\protect\citeauthoryear{Lee, Cheung, and Kao}{Lee
  et~al\mbox{.}}{1998}]%
        {lee1998sampling}
\bibfield{author}{\bibinfo{person}{Sau~Dan Lee}, \bibinfo{person}{David~W
  Cheung}, {and} \bibinfo{person}{Ben Kao}.} \bibinfo{year}{1998}\natexlab{}.
\newblock \showarticletitle{Is sampling useful in data mining? a case in the
  maintenance of discovered association rules}.
\newblock \bibinfo{journal}{\emph{Data Mining and Knowledge Discovery}}
  \bibinfo{volume}{2}, \bibinfo{number}{3} (\bibinfo{year}{1998}),
  \bibinfo{pages}{233--262}.
\newblock


\bibitem[\protect\citeauthoryear{Lin, Alvarez, and Ruiz}{Lin
  et~al\mbox{.}}{2002}]%
        {lin2002efficient}
\bibfield{author}{\bibinfo{person}{Weiyang Lin}, \bibinfo{person}{Sergio~A
  Alvarez}, {and} \bibinfo{person}{Carolina Ruiz}.}
  \bibinfo{year}{2002}\natexlab{}.
\newblock \showarticletitle{Efficient adaptive-support association rule mining
  for recommender systems}.
\newblock \bibinfo{journal}{\emph{Data mining and knowledge discovery}}
  \bibinfo{volume}{6}, \bibinfo{number}{1} (\bibinfo{year}{2002}),
  \bibinfo{pages}{83--105}.
\newblock


\bibitem[\protect\citeauthoryear{Mallik, Mukhopadhyay, and Maulik}{Mallik
  et~al\mbox{.}}{2014}]%
        {mallik2014ranwar}
\bibfield{author}{\bibinfo{person}{Saurav Mallik}, \bibinfo{person}{Anirban
  Mukhopadhyay}, {and} \bibinfo{person}{Ujjwal Maulik}.}
  \bibinfo{year}{2014}\natexlab{}.
\newblock \showarticletitle{RANWAR: rank-based weighted association rule mining
  from gene expression and methylation data}.
\newblock \bibinfo{journal}{\emph{IEEE transactions on nanobioscience}}
  \bibinfo{volume}{14}, \bibinfo{number}{1} (\bibinfo{year}{2014}),
  \bibinfo{pages}{59--66}.
\newblock


\bibitem[\protect\citeauthoryear{Manola, Miller, McBride, et~al\mbox{.}}{Manola
  et~al\mbox{.}}{2004}]%
        {manola2004rdf}
\bibfield{author}{\bibinfo{person}{Frank Manola}, \bibinfo{person}{Eric
  Miller}, \bibinfo{person}{Brian McBride}, {et~al\mbox{.}}}
  \bibinfo{year}{2004}\natexlab{}.
\newblock \showarticletitle{RDF primer}.
\newblock \bibinfo{journal}{\emph{W3C recommendation}} \bibinfo{volume}{10},
  \bibinfo{number}{1-107} (\bibinfo{year}{2004}), \bibinfo{pages}{6}.
\newblock


\bibitem[\protect\citeauthoryear{Meilicke, Chekol, Fink, and
  Stuckenschmidt}{Meilicke et~al\mbox{.}}{2020}]%
        {meilicke2020reinforced}
\bibfield{author}{\bibinfo{person}{Christian Meilicke},
  \bibinfo{person}{Melisachew~Wudage Chekol}, \bibinfo{person}{Manuel Fink},
  {and} \bibinfo{person}{Heiner Stuckenschmidt}.}
  \bibinfo{year}{2020}\natexlab{}.
\newblock \showarticletitle{Reinforced anytime bottom up rule learning for
  knowledge graph completion}.
\newblock \bibinfo{journal}{\emph{arXiv preprint arXiv:2004.04412}}
  (\bibinfo{year}{2020}).
\newblock


\bibitem[\protect\citeauthoryear{Meng and Tu}{Meng and Tu}{2017}]%
        {meng2017flexible}
\bibfield{author}{\bibinfo{person}{Jinghan Meng} {and}
  \bibinfo{person}{Yi-cheng Tu}.} \bibinfo{year}{2017}\natexlab{}.
\newblock \showarticletitle{Flexible and feasible support measures for mining
  frequent patterns in large labeled graphs}. In
  \bibinfo{booktitle}{\emph{SIGMOD}}. \bibinfo{pages}{391--402}.
\newblock


\bibitem[\protect\citeauthoryear{Namaki, Wu, Song, Lin, and Ge}{Namaki
  et~al\mbox{.}}{2017}]%
        {namaki2017discovering}
\bibfield{author}{\bibinfo{person}{Mohammad~Hossein Namaki},
  \bibinfo{person}{Yinghui Wu}, \bibinfo{person}{Qi Song},
  \bibinfo{person}{Peng Lin}, {and} \bibinfo{person}{Tingjian Ge}.}
  \bibinfo{year}{2017}\natexlab{}.
\newblock \showarticletitle{Discovering graph temporal association rules}. In
  \bibinfo{booktitle}{\emph{CIKM}}. \bibinfo{pages}{1697--1706}.
\newblock


\bibitem[\protect\citeauthoryear{Nowozin, Bakir, and Tsuda}{Nowozin
  et~al\mbox{.}}{2007}]%
        {nowozin2007discriminative}
\bibfield{author}{\bibinfo{person}{Sebastian Nowozin}, \bibinfo{person}{Gokhan
  Bakir}, {and} \bibinfo{person}{Koji Tsuda}.} \bibinfo{year}{2007}\natexlab{}.
\newblock \showarticletitle{Discriminative subsequence mining for action
  classification}. In \bibinfo{booktitle}{\emph{ICCV}}. \bibinfo{pages}{1--8}.
\newblock


\bibitem[\protect\citeauthoryear{Ortona, Meduri, and Papotti}{Ortona
  et~al\mbox{.}}{2018}]%
        {ortona2018robust}
\bibfield{author}{\bibinfo{person}{Stefano Ortona},
  \bibinfo{person}{Venkata~Vamsikrishna Meduri}, {and} \bibinfo{person}{Paolo
  Papotti}.} \bibinfo{year}{2018}\natexlab{}.
\newblock \showarticletitle{Robust discovery of positive and negative rules in
  knowledge bases}. In \bibinfo{booktitle}{\emph{ICDE}}.
  \bibinfo{pages}{1168--1179}.
\newblock


\bibitem[\protect\citeauthoryear{Schmitz, Hotho, J{\"a}schke, and
  Stumme}{Schmitz et~al\mbox{.}}{2006}]%
        {schmitz2006mining}
\bibfield{author}{\bibinfo{person}{Christoph Schmitz}, \bibinfo{person}{Andreas
  Hotho}, \bibinfo{person}{Robert J{\"a}schke}, {and} \bibinfo{person}{Gerd
  Stumme}.} \bibinfo{year}{2006}\natexlab{}.
\newblock \showarticletitle{Mining association rules in folksonomies}.
\newblock In \bibinfo{booktitle}{\emph{Data science and classification}}.
  \bibinfo{pages}{261--270}.
\newblock


\bibitem[\protect\citeauthoryear{Shelokar, Quirin, and Cord{\'o}n}{Shelokar
  et~al\mbox{.}}{2014}]%
        {shelokar2014three}
\bibfield{author}{\bibinfo{person}{Prakash Shelokar}, \bibinfo{person}{Arnaud
  Quirin}, {and} \bibinfo{person}{{\'O}scar Cord{\'o}n}.}
  \bibinfo{year}{2014}\natexlab{}.
\newblock \showarticletitle{Three-objective subgraph mining using
  multiobjective evolutionary programming}.
\newblock \bibinfo{journal}{\emph{J. Comput. System Sci.}}
  \bibinfo{volume}{80}, \bibinfo{number}{1} (\bibinfo{year}{2014}),
  \bibinfo{pages}{16--26}.
\newblock


\bibitem[\protect\citeauthoryear{Thompson}{Thompson}{2002}]%
        {thompson2002sampling}
\bibfield{author}{\bibinfo{person}{S Thompson}.}
  \bibinfo{year}{2002}\natexlab{}.
\newblock \bibinfo{booktitle}{\emph{Sampling}}.
\newblock \bibinfo{publisher}{Wiley}.
\newblock


\bibitem[\protect\citeauthoryear{Toivonen et~al\mbox{.}}{Toivonen
  et~al\mbox{.}}{1996}]%
        {toivonen1996sampling}
\bibfield{author}{\bibinfo{person}{Hannu Toivonen} {et~al\mbox{.}}}
  \bibinfo{year}{1996}\natexlab{}.
\newblock \showarticletitle{Sampling large databases for association rules}. In
  \bibinfo{booktitle}{\emph{VLDB}}, Vol.~\bibinfo{volume}{96}.
  \bibinfo{pages}{134--145}.
\newblock


\bibitem[\protect\citeauthoryear{Vanetik, Gudes, and Shimony}{Vanetik
  et~al\mbox{.}}{2002}]%
        {vanetik2002computing}
\bibfield{author}{\bibinfo{person}{Natalia Vanetik}, \bibinfo{person}{Ehud
  Gudes}, {and} \bibinfo{person}{Solomon~Eyal Shimony}.}
  \bibinfo{year}{2002}\natexlab{}.
\newblock \showarticletitle{Computing frequent graph patterns from
  semistructured data}. In \bibinfo{booktitle}{\emph{ICDM}}.
  \bibinfo{pages}{458--465}.
\newblock


\bibitem[\protect\citeauthoryear{Vrande{\v{c}}i{\'c} and
  Kr{\"o}tzsch}{Vrande{\v{c}}i{\'c} and Kr{\"o}tzsch}{2014}]%
        {vrandevcic2014wikidata}
\bibfield{author}{\bibinfo{person}{Denny Vrande{\v{c}}i{\'c}} {and}
  \bibinfo{person}{Markus Kr{\"o}tzsch}.} \bibinfo{year}{2014}\natexlab{}.
\newblock \showarticletitle{Wikidata: a free collaborative knowledgebase}.
\newblock \bibinfo{journal}{\emph{Commun. ACM}} \bibinfo{volume}{57},
  \bibinfo{number}{10} (\bibinfo{year}{2014}), \bibinfo{pages}{78--85}.
\newblock


\bibitem[\protect\citeauthoryear{Wang, Xu, and Zhan}{Wang
  et~al\mbox{.}}{2020}]%
        {wang2020extending}
\bibfield{author}{\bibinfo{person}{Xin Wang}, \bibinfo{person}{Yang Xu}, {and}
  \bibinfo{person}{Huayi Zhan}.} \bibinfo{year}{2020}\natexlab{}.
\newblock \showarticletitle{Extending association rules with graph patterns}.
\newblock \bibinfo{journal}{\emph{Expert Systems with Applications}}
  \bibinfo{volume}{141} (\bibinfo{year}{2020}), \bibinfo{pages}{112897}.
\newblock


\bibitem[\protect\citeauthoryear{Yan and Han}{Yan and Han}{2002}]%
        {yan2002gspan}
\bibfield{author}{\bibinfo{person}{Xifeng Yan} {and} \bibinfo{person}{Jiawei
  Han}.} \bibinfo{year}{2002}\natexlab{}.
\newblock \showarticletitle{gspan: Graph-based substructure pattern mining}. In
  \bibinfo{booktitle}{\emph{ICDM}}. \bibinfo{pages}{721--724}.
\newblock


\bibitem[\protect\citeauthoryear{Yang}{Yang}{2004}]%
        {yang2004complexity}
\bibfield{author}{\bibinfo{person}{Guizhen Yang}.}
  \bibinfo{year}{2004}\natexlab{}.
\newblock \showarticletitle{The complexity of mining maximal frequent itemsets
  and maximal frequent patterns}. In \bibinfo{booktitle}{\emph{KDD}}.
  \bibinfo{pages}{344--353}.
\newblock


\bibitem[\protect\citeauthoryear{Zaki, Lesh, and Ogihara}{Zaki
  et~al\mbox{.}}{1998}]%
        {zaki1998planmine}
\bibfield{author}{\bibinfo{person}{Mohammed~Javeed Zaki}, \bibinfo{person}{Neal
  Lesh}, {and} \bibinfo{person}{Mitsunori Ogihara}.}
  \bibinfo{year}{1998}\natexlab{}.
\newblock \showarticletitle{PlanMine: Sequence Mining for Plan Failures.}. In
  \bibinfo{booktitle}{\emph{KDD}}. \bibinfo{pages}{369--373}.
\newblock


\bibitem[\protect\citeauthoryear{Zhao and Bhowmick}{Zhao and Bhowmick}{2003}]%
        {zhao2003association}
\bibfield{author}{\bibinfo{person}{Qiankun Zhao} {and}
  \bibinfo{person}{Sourav~S Bhowmick}.} \bibinfo{year}{2003}\natexlab{}.
\newblock \showarticletitle{Association rule mining: A survey}.
\newblock \bibinfo{journal}{\emph{Nanyang Technological University}}
  \bibinfo{volume}{135} (\bibinfo{year}{2003}).
\newblock


\end{thebibliography}






\end{document}